\documentclass[aps,pre,twocolumn,letterpaper,floatfix,showpacs]{revtex4}
\usepackage{graphicx} 
\usepackage{amsmath,amssymb,amsfonts} 
\usepackage{mathtools}
\usepackage[hidelinks]{hyperref} 
\usepackage{epstopdf}
\newcommand{\smax}{{s_\mathrm{m}}}

\newcommand{\ds}{\mathrm{d}s}
\newcommand{\dt}{\mathrm{d}t}
\newcommand{\dx}{\mathrm{d}x}
\newcommand{\fNmeso}{{f_{w}}}

\newcommand{\ta}{{t_a}}
\newcommand{\dta}{{\mathrm{d}t_a}}
\newcommand{\A}{A}
\newcommand{\PS}{\frac{\smax}{\smax + v\langle t_a \rangle}}
\newcommand{\PA}{\frac{v\langle t_a\rangle}{\smax + v\langle t_a \rangle}}
\newcommand{\firstderiv}[2]{\frac{\mathrm{d}{#1}}{\mathrm{d}{#2}}}
\newcommand{\BraunMEanalytic}{Braun2008modeling,Braun2010master}
\newcommand{\BraunNoTimeDependence}{Braun2008modeling,Braun2010master}
\begin{document}
\title{History-dependent friction and slow slip from time-dependent microscopic junction laws studied in a statistical framework}
\author{Kjetil Th{\o}gersen}
\email{kjetil.thogersen@fys.uio.no}
\author{J{\o}rgen Kjoshagen Tr{\o}mborg}
\author{Henrik Andersen Sveinsson}
\author{Anders Malthe-S{\o}renssen}
\affiliation{Department of Physics\\University of Oslo\\Sem S{\ae}lands vei 24, \\ NO-0316, Oslo, Norway}
\author{Julien Scheibert}
\affiliation{Laboratoire de Tribologie et Dynamique des Syst{\'e}mes, CNRS, Ecole Centrale de Lyon\\36, Avenue Guy de Collongue, \\ 69134 Ecully CEDEX, France}
\date{\today} 

\begin{abstract}
To study how macroscopic friction phenomena originate from microscopic junction laws, we introduce a general statistical framework describing the collective behavior of a large number of individual micro-junctions forming a macroscopic frictional interface. Each micro-junction can switch in time between two states: A pinned state characterized by a displacement-dependent force, and a slipping state characterized by a time-dependent force. Instead of tracking each micro-junction individually, the state of the interface is described by two coupled distributions for (i) the stretching of pinned junctions and (ii) the time spent in the slipping state. This framework allows for a whole family of micro-junction behavior laws, and we show how it represents an overarching structure for many existing models found in the friction literature. We then use this framework to pinpoint the effects of the time-scale that controls the duration of the slipping state. First, we show that the model reproduces a series of friction phenomena already observed experimentally. The macroscopic steady-state friction force is velocity-dependent, either monotonic (strengthening or weakening) or non-monotonic (weakening-strengthening), depending on the microscopic behavior of individual junctions. In addition, slow slip, which has been reported in a wide variety of systems, spontaneously occurs in the model if the friction contribution from junctions in the slipping state is time-weakening. Next, we show that the model predicts a non-trivial history-dependence of the macroscopic static friction force. In particular, the static friction coefficient at the onset of sliding is shown to increase with increasing deceleration during the final phases of the preceding sliding event. We suggest that this form of history-dependence of static friction should be investigated in experiments, and we provide the acceleration range in which this effect is expected to be experimentally observable.

\end{abstract}

\maketitle

\section{Introduction}

Solid friction is of considerable importance to a large number of fields, from geological \cite{Scholz-CUP-2002} to biological \cite{Scheibert-Leurent-Prevost-Debregeas-Science-2009}, engineering \cite{zoback2012importance,Das2011long-period} and materials \cite{Bartlett2012looking} sciences. It originates at the microscopic scales of the interface between two solids in contact. However, problems in friction often couple various time and length scales \cite{persson2000sliding, Baumberger2006solid, vanossi2013colloquium}. To describe friction at large scales, upscaled/macroscopic friction laws are needed. Such laws are commonly formulated on length scales at which the local structure of the interface is assumed to be averaged out. The Amontons--Coulomb laws \cite{Coulomb-Bachelier-1821}, the rate and state laws \cite{Rice-Ruina-JApplMech-1983, BarSinai2012slow} and other macroscopic friction laws parametrize the frictional response of the interface when submitted to external forces in terms of a handful of friction parameters, e.g. the static and kinetic friction coefficients. The microscopic origin of the friction forces does not explicitly enter in these descriptions, but is usually invoked to justify the basic features of the laws chosen for a given system, e.g. a proportionality between friction and normal forces.

The microscopic forces responsible for friction vary between systems. They can for example be associated with micro-contacts between asperities in rough interfaces \cite{Greenwood1966contact,Baumberger2006solid}, pinned islands in boundary lubrication \cite{Persson1995theory}, or molecular bonds \cite{Schallamach1963theory,Filippov2004friction,Srinivasan2009binding}. We use the term micro-junction to refer to a single micro-contact, island or bond. To create a fundamental description of friction that takes its microscopic origins explicitly into account, two questions must be answered: (i) What is the behavior law for a given micro-junction? (ii) How can we upscale/integrate these laws to deduce the friction behavior at a larger length scale involving a large number of micro-junctions? The first question is addressed by the field of nanotribology (see e.g. \cite{szlufarska2008recent}). Here we address the second question. In particular, we investigate the consequences that a time-dependent micro-junction behavior law has on the macroscopic friction force.

In principle, the state of a multi-junction interface could be monitored by following the individual state of each micro-junction. In practice, this task may not be possible for a series of reasons. First, the number of micro-junctions can be large, making it difficult to keep track of all the time evolutions of the parameters defining their individual states. Second, the properties of individual junctions (e.g. size, stiffness or threshold) are often known only in a statistical sense. Third, the external forces/stresses on the junctions are only known in average, through the total macroscopic applied loads on the whole interface.

A way around the above mentioned difficulties is the following: Instead of tracking individual junctions as they are loaded or start to slip under a small additional strain, the fraction of junctions that are loaded or start to slip is monitored.  The idea of considering distributions rather than a finite set of micro-junctions will be used extensively here, and has also been used previously in various studies of friction (e.g. \cite{Schallamach1963theory, Persson1995theory, Farkas2005static, Braun2008modeling, Braun2010master}). Farkas et al. \cite{Farkas2005static} studied the evolution of the junctions' friction forces as a function of the displacement of a rigid slider. In particular, they showed how the macroscopic friction force depends on the distributions of both the shear stresses and strengths among the population of individual micro-junctions. Recently, Braun and Peyrard \cite{Braun2008modeling} showed that the evolution of the friction force as a function of the displacement of the slider can be solved with a differential equation -- the master equation. Using this framework, they could study the relationship between the distribution of junction strengths and the occurence of either stick-slip motion or smooth sliding \cite{Braun2010master}.

In these studies, the friction force was displacement dependent only. However, there is overwhelming experimental evidence that friction does not depend only on displacement. Among other phenomena: most interfaces have a velocity-dependent steady-state friction behavior (see e.g. \cite{Grosch1963relation, Baumberger2006solid, BenAbdelounis2010noise}); most interfaces are aging, i.e. have a strength that increases with increasing time spent in contact before slip (see e.g. \cite{Dieterich1978Time, Baumberger2006solid, Li2011ageing}); the slip dynamics at short times after slip inception in polymethylmethacrylate is controlled by a time scale \cite{Ben-David2010slip-stick}; the healing rate of seismic faults after an earthquake varies after a characteristic time scale \cite{Marone1998effect}; the friction force during reciprocating/oscillating motion depends on whether slip is accelerating or decelerating \cite{Bureau2001lowvelocity}.

Motivated by these observations, a number of models have introduced time dependencies in the behavior of individual micro-junctions. Within the distribution approach, Schallamach introduced time rates for both the thermally activated bonding and de-bonding of molecules onto a surface to model the velocity-dependent friction of rubber \cite{Schallamach1963theory};  Persson, in a study of contacts with a lubrication film of molecular thickness (boundary lubrication), introduced a similar rate, but for the bonding of pinned adsorbate domains only \cite{Persson1995theory}; Braun and Peyrard also considered, in the master equation framework, the effects of a constant time delay for the repinning of micro-junctions and of an increase in the strength with the age of a pinned junction \cite{Braun2011dependence}. Numerically, time-scales were also introduced for finite sets of micro-junctions put in parallel to model the friction between a surface and a slider. Filippov et al. used bonding and de-bonding time rates to model adhesive boundary lubricated surfaces and cold welding \cite{Filippov2004friction}; in order to study micro-slip front propagation at a frictional interface, various models recently considered elastic sliders made of blocks connected by internal springs, each block being itself connected to the surface by a series of micro-junctions \cite{Braun2009dynamics,Capozza2011stabilizing,Capozza2012static,Tromborg2014slow}. Realistic results could be obtained using time delays between depinning and repinning of junctions.

Here, we present a general framework for models in which micro-junctions can switch between a time-dependent and a displacement-dependent state (Section \ref{sec:definingTheModelAndTheDistributions}). The framework provides an explicit description of the distributions of individual junction states. It also allows for analytical continuum predictions that are useful to provide a systematic understanding of the effects of time-dependent junction laws. The framework can be applied to a whole family of behavior laws at the microscopic scale, and we show how previously studied models \cite{Farkas2005static, Braun2008modeling,Braun2010master, Braun2011dependence, Braun2009dynamics, Capozza2011stabilizing,Capozza2012static, Tromborg2014slow} are subsets of the general framework. We then explore the macroscopic consequences of microscopic variability in the transition time between the time-dependent and the displacement-dependent states. We first derive a general expression for the steady state friction force and show how it is directly related to the microscopic junction behavior. Depending on the microscopic laws used, the model can exhibit monotonic (strengthening or weakening) as well as non-monotonic velocity dependencies, all of which have been observed experimentally \cite{Chen2006velocity,bar2013velocity,Dieterich1978Time, Heslot1994creep, Baumberger2006solid} (Section \ref{sec:Steady state}). In addition to this steady state phenomenology, the model gives insight into transient phenomena and static friction (Section \ref{sec:transients}). In particular, we show how the static friction coefficient is directly related to the distribution of shear forces on individual micro-junctions. We also predict a non-trivial history-dependence of the macroscopic static friction coefficient at the onset of sliding. More precisely, it is strongly influenced by the deceleration dynamics of the final phases of the  preceding sliding event. We also show that slow slip, which has been reported in a wide variety of systems (see e.g. \cite{Ohnaka1990constitutive, Heslot1994creep, Hirose2005repeating, Peng2010integrated, Ben-David2010slip-stick,Yang2008dynamics}), spontaneously occurs in the framework if the friction force contribution from junctions in the slipping state is time-weakening. Section \ref{sec:summary_and_conclusion} contains the discussion and conclusion.

\section{Model description\label{sec:definingTheModelAndTheDistributions}}
\begin{figure}
\centering
\includegraphics[width=.49\textwidth]{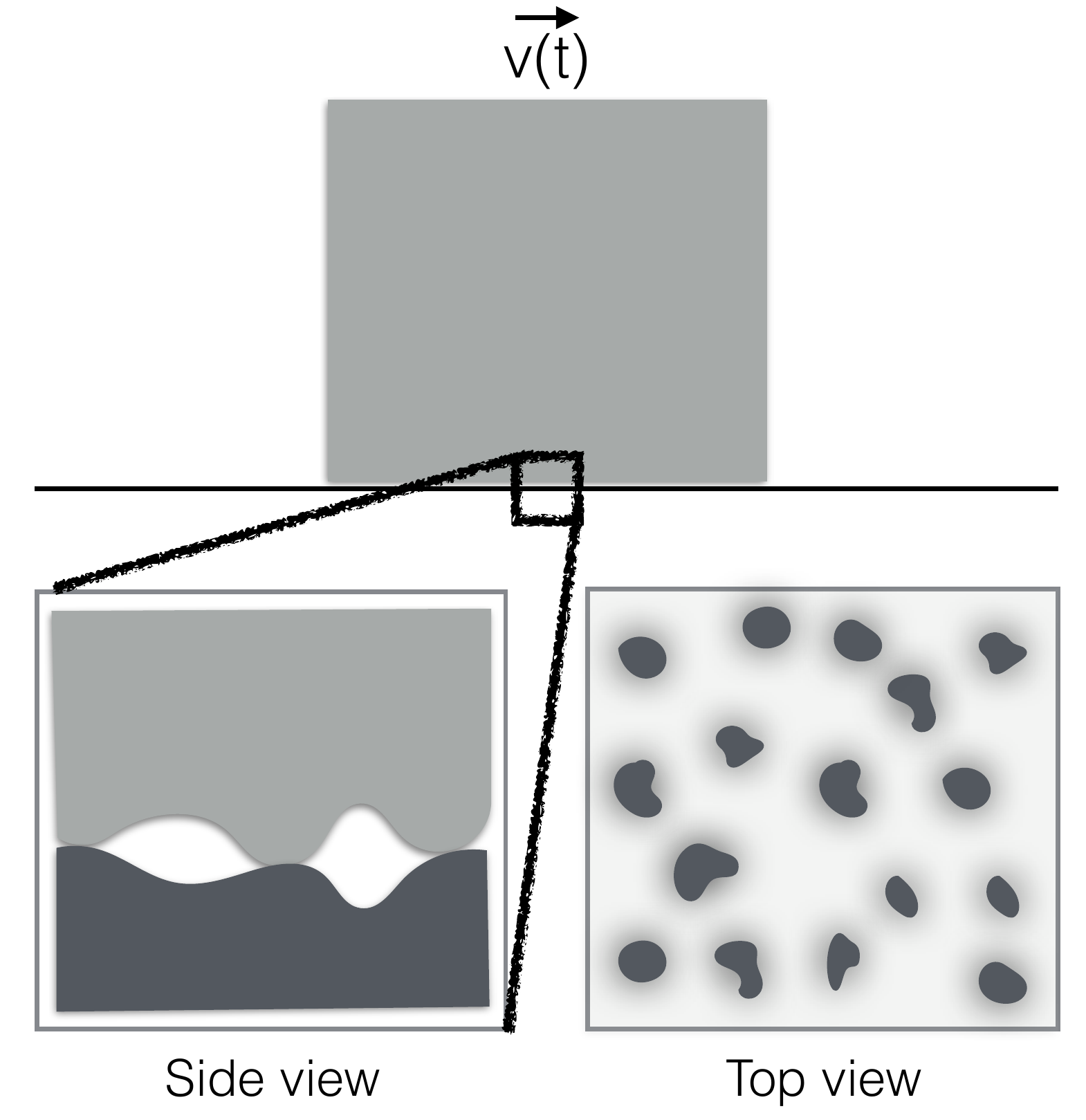}
\caption{Sketch of the system we model. We consider a nominally flat frictional contact between a moving rigid slider and a track (top). The interface consists of a large number of individual micro-junctions. For rough solids, the junctions correspond to micro-contacts between antagonist asperities (bottom left), which are distributed spatially across the apparent contact area (bottom right).
\label{fig:system_sketch}}
\end{figure}

We study the frictional behavior of a rigid slider (macroscopic block) that interacts with its substrate through a large number of 
micro-junctions (\figurename~\ref{fig:system_sketch}). The junctions are assumed to be independent. They are all stretched by equal amounts when the slider moves. This assumption is valid if the lateral size of the slider is smaller than the elastic screening length, $\xi$, so that the interface can be considered rigid \cite{Caroli1998hysteresis, Braun2012collective}. To study systems that are larger than $\xi$, elastic interactions must be accounted for, for example by using spring--block models, with blocks of size $\xi$, as in \cite{Braun2009dynamics, Tromborg2011transition, Amundsen20121D,Capozza2011stabilizing,Capozza2012static,Tromborg2014slow}.

\subsection{The behaviour of individual junctions}
\begin{figure}
\centering
\includegraphics[width=.35\textwidth]{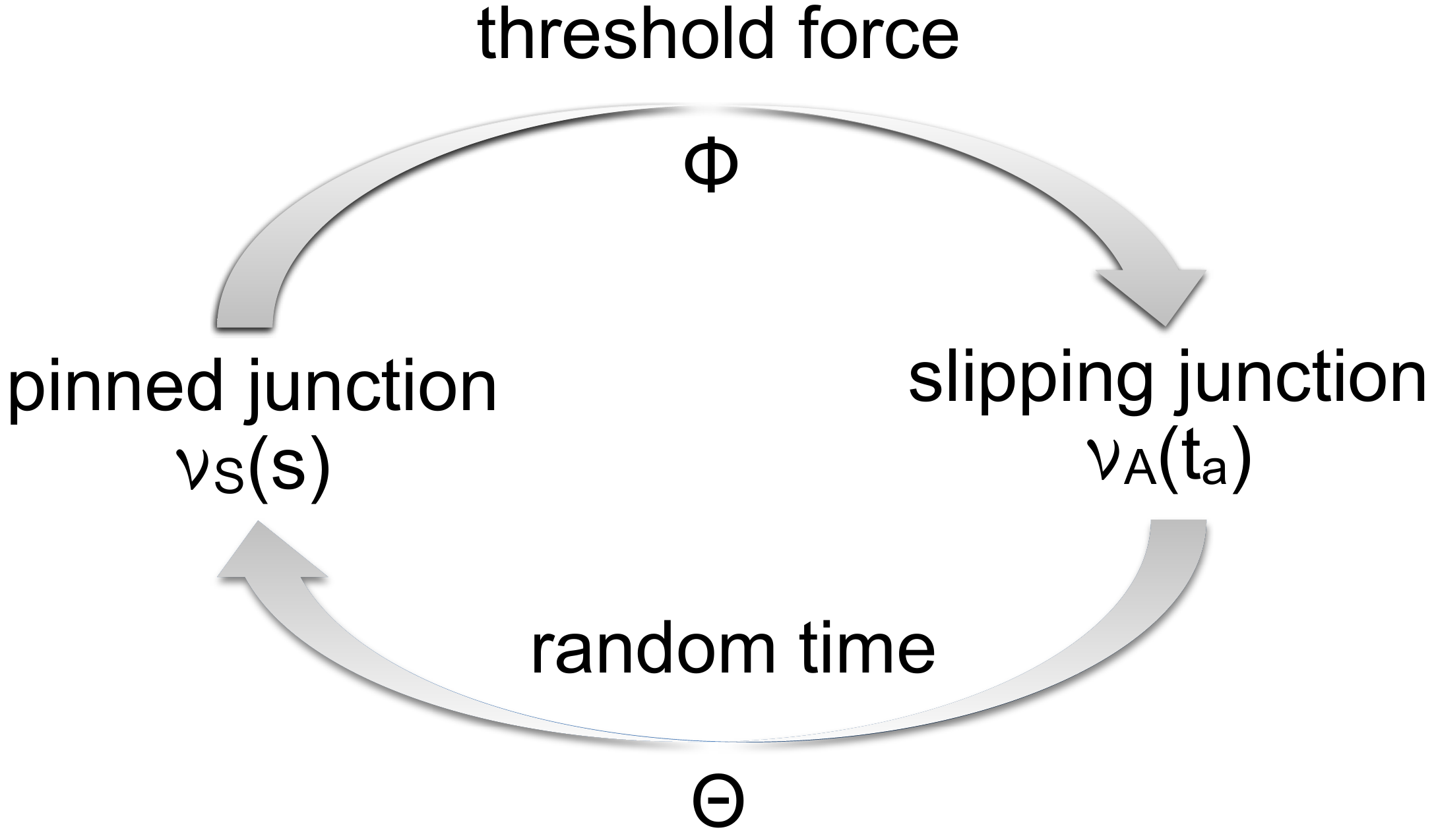}
\caption{Junctions can exist in two different states, the pinned state and the slipping state. The transition from the pinned state to the slipping state is governed by a force threshold, described by the function $\Phi(s)$, where $s$ is the junction's stretching. The transition from the slipping state to the pinned state occurs after a random time (also called delay time) controlled by the function $\Theta(t_a)$, where $t_a$ is the time spent in the slipping state. The force in the pinned state is a function of the junction stretching, and is given by $f_S(s)$. The force in the slipping state is a function of the time spent in the slipping state, and is given by $f_A(t_a)$. The dimensionless versions of these forces are $\nu_S(s)  = Nf_S/\fNmeso$ and $\nu_A(t_a) = Nf_A(t_a)/\fNmeso$, where $N$ is the number of junctions and $\fNmeso$ is the normal force.
\label{fig:one_junction_sketch}}
\end{figure}

We assume that individual junctions can exist in a displacement-dependent state, the pinned state; and in a time-dependent state, the slipping state. The junctions switch states as sketched in \figurename~\ref{fig:one_junction_sketch}. A junction remains in the pinned state until it is stretched beyond its breaking threshold force. It then enters the slipping state, where it stays for a random time (also called delay time), after which it is repinned or replaced by a different junction. In general, the force from each junction on the slider depends on the state of the junction. A pinned junction acts with a force $f_\text{pinned} = f_S(s)$, where $s$ is the stretching (the distance from the pinning point of the junction at the substrate to its attachment point on the moving slider). A slipping junction contributes a force that can depend on the time spent in the slipping state $\ta$, $f_\text{slipping} = f_A(\ta)$. $f_S$ and $f_A$ are not necessarily displacement and time-dependent only; they can depend on other physical quantities, such as temperature and the velocity of the slider. 

\subsection{A general framework for collective junction behaviour}
To study the macroscopic friction force we need to know the collective behavior of a large number of junctions. In this section we introduce a general framework for the collective junction behavior, and show how various recent models are subsets of the framework. We then reduce the number of parameters and study the effect of disorder in the time at which the slipping-to-pinned transition occurs.

\subsubsection{Junction state distributions}
When the number of junctions is large there is no need to keep track of the state of each individual junction. Instead, the collective state of the junctions can be described by two probability densities; one holding the information about pinned junctions and another holding the information about slipping junctions. Knowledge of these distributions can be used to determine the main variable of interest: The macroscopic friction force. In general, the instantaneous values of the distributions will depend on the past and present slip history of the slider.

Consider a system of $N$ junctions. Every time a junction leaves one of the states, the junction or its replacement enters the other state, so the total number stays unchanged.
\begin{align}
\begin{split}
N &= \text{number of pinned junctions}\\
    &+\text{number of slipping junctions}.
    \label{eq:sum_normalization}
\end{split}
\end{align}
 This normalization condition can be written in a continuum formulation as
\begin{align}
1 &= \int_{-\infty}^{\infty} S(s)\, \ds + \int_0^{\infty} \A(\ta) \,\dta,\qquad\forall t.\label{eq:integral_normalization}
\end{align}
Equations~\eqref{eq:sum_normalization} and \eqref{eq:integral_normalization} differ by a factor $N$ which will be absorbed into the force law. The stretching probability density $S(s)$ holds the information about the stretching of pinned junctions. The slipping time probability density $A(\ta)$ holds the information about the slipping time of slipping junctions. These distributions evolve with global time $t$ and with the motion of the slider, $x(t)$, so that $S = S(s,t)$ and $A = A(\ta,t)$. The two time variables $\ta$ and $t$ evolve with the same increments, but serve different roles in the formalism. The global time $t$ is used to determine chronology and simultaneity, so that $x(t)$ and $S(s,t)$ are values taken at the same point in time. The slipping time $\ta$, on the other hand, takes on different values for different junctions, or in the integral formulation, for different parts of $A$; because junctions enter the slipping state at different instants in time.

The macroscopic friction force $f_\text{macro}$ on the slider is the sum of the forces from all junctions. The contribution to $f_\text{macro}$ from the pinned junctions is a function of their stretching, $s$, and the contribution from the slipping junctions in general depends on the slipping time, $t_a$. We have that
\begin{align}
f_\text{macro}
  &= \sum_{i=1}^{\text{pinned junctions}} f_S(s_i) + \sum_{i=1}^{\text{slipping junctions}} f_A(t_{a,i})\label{eq:sum_fmeso}.
\end{align}
The corresponding equation in the integral formulation is 
\begin{align}
\nu_\text{macro}
  &\equiv f_\text{macro}/\fNmeso\\
  &= \int_{-\infty}^\infty \nu_S(s)S(s)\,\ds + \int_0^\infty \nu_A(\ta)\A(\ta)\,\dta, \label{eq:total_friction}
\end{align}
where $\fNmeso$ is the normal force on the slider and $\nu_S \equiv Nf_S/\fNmeso$, $\nu_A \equiv Nf_A/\fNmeso$. Two comments are in order at this point. First, we have absorbed the $N$ in the force law in order to have $S$ and $A$ normalized to $1$. Second, the calculations that follow in the rest of the paper benefit from using a non-dimensional formulation of the force law and so we have divided by the normal force. Independently of wether the friction forces are proportional to the normal force, or have some other normal force dependence, this is the characteristic force level in the system. We will not study the effect of a varying normal force in this paper, so $f_\text{macro}$ can always be recovered from $\nu_\text{macro}$ by multiplying with $\fNmeso$.

\subsubsection{Evolution of S and A with time and displacement\label{sec:evolutionOfSAndA}}
\begin{figure}
\centering
\includegraphics[width=.49\textwidth]{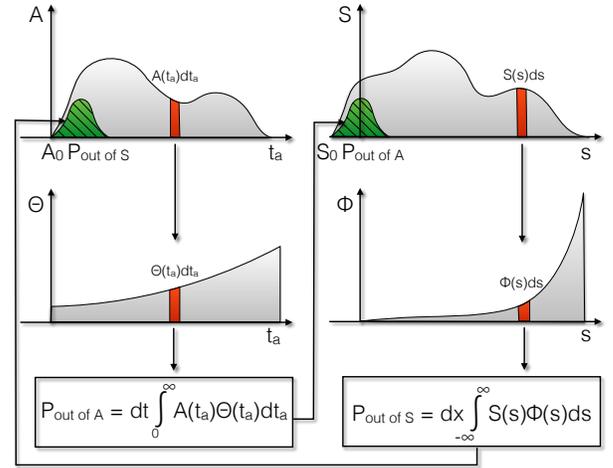}
\caption{(Color online) Evolution cycle for the distributions of pinned and slipping junctions. 
\emph{Top, left}: Distribution of junctions in the slipping state, $A$. The probability to find a junction with slipping time $\in[\ta,\ta+\dta]$ is $A(t_a)\dt_a$ (\emph{red (dark gray)}). The junctions that enter the slipping state from $S$, add to the slipping time distribution as $A_0 P_\text{out of S}$ (\emph{green (dark gray) hatched}). \emph{Middle, left}: The probability of leaving the slipping state is governed by the distribution $\Theta$.
\emph{Bottom, left}: The fraction of junctions that leave $A$ during an infinitesimal time step $\dt$ is an integral over $A$ multiplied by $\Theta$.
\emph{Top, right}: Distribution of junctions in the pinned state, $S$. The junctions that enter $S$ are assigned an initial stretching distribution $S_0$. This probability is added to $S$ as $S_0 P_\text{out of A}$ (\emph{green (dark gray) hatched}). The probability to find a pinned junction at stretching $\in[s,s+\ds]$ is $S(s) \ds$ (\emph{red (dark gray)}).
\emph{Middle, right}: The probability that leaves $S$ during a displacement $\dx$ is given by the distribution $\Phi$.
\emph{Bottom right}: The fraction of junctions that leave $S$ during an infinitesimal displacement $\dx$ is an integral over $S$ multiplied by $\Phi$. This probability enters $A$, and the cycle is complete.
\label{fig:general_model_description}} 
\end{figure}

The equations in the previous section apply when $S(s)$ and $\A(\ta)$ are known. To use them, we also need to know how the distributions develop in time as the slider moves. \figurename~\ref{fig:general_model_description} shows the cycle during an infinitesimal time interval $\dt$. The changes in the distributions have three contributions: 1) Rigid shift of the distribution. 2) Moving probability out of the distribution (junctions that break or repined/reform). 3) Probability received from the other distribution.

We start with the probability that moves from $S(s)$ to $A(t_a)$. Recall that $S$ does not explicitly include a time dependence, but changes due to the changes in the position of the slider, $x(t)$. If the slider moved the distance $\dx$ during the infinitesimal time interval $\dt$, the fraction of contacts that broke is
\begin{align}
P_\text{out of S} = \dx\, \int_{-\infty}^\infty S(s,t) \Phi(s)\, \ds,
\label{eq:P_out_of_S}
\end{align}
where $\Phi(s)\ds$ is the probability for a contact with a stretching $s$ to break during a mesoscopic displacement $\dx$. In Appendix~\ref{appsec:Phi_equiv_smaxDistribution} we show that this formulation is mathematically equivalent to having a distribution of junction strengths/thresholds, and how this function can be derived from $\Phi$.

In a formally equivalent way, the probability that moves from $A(t_a)$ to $S(s)$ (the fraction of junctions that repin/reform during a time interval $\dt$) is given by an integral over the distribution of slipping times multiplied by a probability function $\Theta(\ta)$,
\begin{align}
P_\text{out of \A} = \dt\, \int_0^{\infty} \A(\ta,t) \Theta(\ta)\, \dta.\label{eq:PoutofAdefined}
\end{align}
Here we have written $\Theta$ in a simple form where it only depends on the time since slipping was initiated $t_a$, but it could also depend on other parameters such as the velocity of the slider, temperature, and so on. In this way more complicated rules for the evolution of slipping junctions can be modeled. In appendix~\ref{appsec:Theta_equiv_tau} we show that $\Theta$ is mathematically equivalent to having a distribution of delay times.

Note that while $S$ and $A$ have the normalization condition in equation~\eqref{eq:integral_normalization}, $\Phi$ and $\Theta$ can integrate to arbitrary values. For example, we will later use $\Theta=\text{constant}$ for all $\ta$, whose integral over all $\ta$ diverges.

The portion of contacts that enter the pinned state will be assigned a stretching given by an initial distribution of stretchings $S_0$. Combining the terms for contacts that leave and enter the pinned state with the rigid shift of $S$, the evolution rule becomes
\begin{align}
\begin{split}
S(s,t+\dt)  = &  S(s-\dx,t)\left (1 - \dx \,\Phi(s) \right ) \\
& + S_0(s) P_\text{out of A}.
\label{eq:S_evolution}
\end{split}
\end{align}

A formal equivalence for $A$ is achieved if contacts that break are assigned a non-zero initial slipping time from a distribution $A_0$,
\begin{align}
\begin{split}
\A(\ta,t+\dt)  = &  \A(\ta-\dt,t)\left (1 - \dt \,\Theta(\ta) \right )\\
& + \A_0(\ta) P_\text{out of S}.
\end{split}
\end{align}
$A_0$ should intuitively be a $\delta$-function at $\ta=0$, but we state it here because it gives symmetry to the equations and could in principle give more possibilities for the force law in the time-dependent state.

\subsection{Relating the general framework to previously published friction models\label{sec:RelationToPreviousModels}}
We are not the first to study friction models that fit in the framework defined above. In fact, our framework can be seen as a natural extension of existing results. In this section we give a few examples of previously studied models that are subsets of the general framework.

Farkas et al. \cite{Farkas2005static} studied a model where junctions are represented by linear elastic springs. In the general framework this translates into $f_S(s) = ks$, where $k$ is the shear stiffness of the junction. They investigated the influence of the distribution of junction stretchings on macroscopic friction force. The source of disorder in their model is that the junctions have different breaking thresholds. In the general framework this is encoded in $\Phi$, which can be mapped directly to a distribution of contact strengths (Appendix~\ref{appsec:Phi_equiv_smaxDistribution}). Whenever one junction reaches its breaking threshold (strengths), it is immediately replaced by an unloaded junction with the same properties. This is the special case of letting $S_0 = \delta(s-0)$, $A_0=\delta(\ta-0)$, $\Theta(\ta) = \delta(\ta-0)$.

Braun and Peyrard \citep{\BraunMEanalytic} studied a model similar to the one of Farkas et al. They developed a master equation formulation of their model and used it to study the stick-slip and steady sliding regimes and the dependence of kinetic friction on sliding velocity. In \cite{\BraunNoTimeDependence} they assume that there is no time-dependent state of the junctions, that is, junctions that reach their breaking threshold are immediately replaced by new junctions pinned at a lower stretching. This is a subset of the general framework that is realized by letting $A_0=\delta(t_a - 0)$, $\Theta(\ta) = \delta(\ta-0)$. Further, they introduce the forces in pinned springs as $f_S(s_i) = k_is_i(t)$, that is, with an individual spring stiffness, but in the actual calculations they use $f_S(s_i) = \langle k_i\rangle s_i(t)$, with $\langle k_i\rangle$ the average spring stiffness. In place of $\Phi(s)$ they use a distribution of spring stretching thresholds $P_c(s)$; Appendix~\ref{appsec:Phi_equiv_smaxDistribution} gives the mapping between these formulations. Their distribution $R$ of the initial spring stretchings is equivalent to $S_0$ in the general framework.

In \cite{Farkas2005static,\BraunNoTimeDependence} the evolution of the junction states is controlled by a single variable, the position $x(t)$ of the rigid slider. As a consequence of the disorder in breaking thresholds, these systems always approach a steady state when the block displacement is sufficiently large. 

Braun and other co-workers introduced an enriched model in which junctions that reach their breaking threshold are removed and replaced by new unloaded junctions after a delay time $\tau$ \cite{Braun2009dynamics}. As for $P_c$ and $\Phi$ there is a mapping between $\tau$ and $\Theta$ (it has the same mathematical form and can be found in appendix~\ref{appsec:Theta_equiv_tau}). This model is the special case of $S_0=\delta(0)$, $A_0=\delta(0)$, $\Theta = \delta(t_a-\tau)$ and $f_A = 0$. There is also a viscous force term in their model, that acts directly on the slider. This could be added as an additional term in equation \eqref{eq:total_friction}. The model formulated by \cite{Braun2009dynamics} has also been used by Capozza et al. \cite{Capozza2011stabilizing, Capozza2012static}. In Ref.~\cite{Capozza2011stabilizing} they also use a distribution of delay times, which would correspond to defining a $\Theta \neq \delta(t_a-\tau)$.

The model used by us in \cite{Tromborg2014slow} can be formulated in the general framework by using the mapping between $\Theta$ and a distribution of time that junctions will remain in the slipping state $\tau(t_r)$. The additional assumptions are $f_S(s) = ks$, $f_A = \text{constant}$, $S_0 = \delta(s-0)$, $\Phi = \delta(s-\smax)$.

Following up on the idea of a delay time, Braun and Peyrard did a combined study of temperature activated breaking of pinned junctions, an increase of junction strength with time (ageing) and a delay time \cite{Braun2011dependence}. The major component included in our framework, but not included in \cite{Braun2011dependence}, is a disorder in the time spent in the slipping state. This is the focus of the present article. Although we have not included temperature effects and ageing in our presentation, the approach taken by \cite{Braun2011dependence} could be reused in our general framework: thermal breaking could be added as an additional time-and-temperature-dependent term in equations \eqref{eq:P_out_of_S} and \eqref{eq:S_evolution}, and ageing could be included through a time-dependence in $\Phi$.

\subsection{Simplified model: Time-dependent junction behavior \label{sec:time_dependent_junction_behavior}}
\begin{figure}
\centering
\includegraphics[width=.49\textwidth]{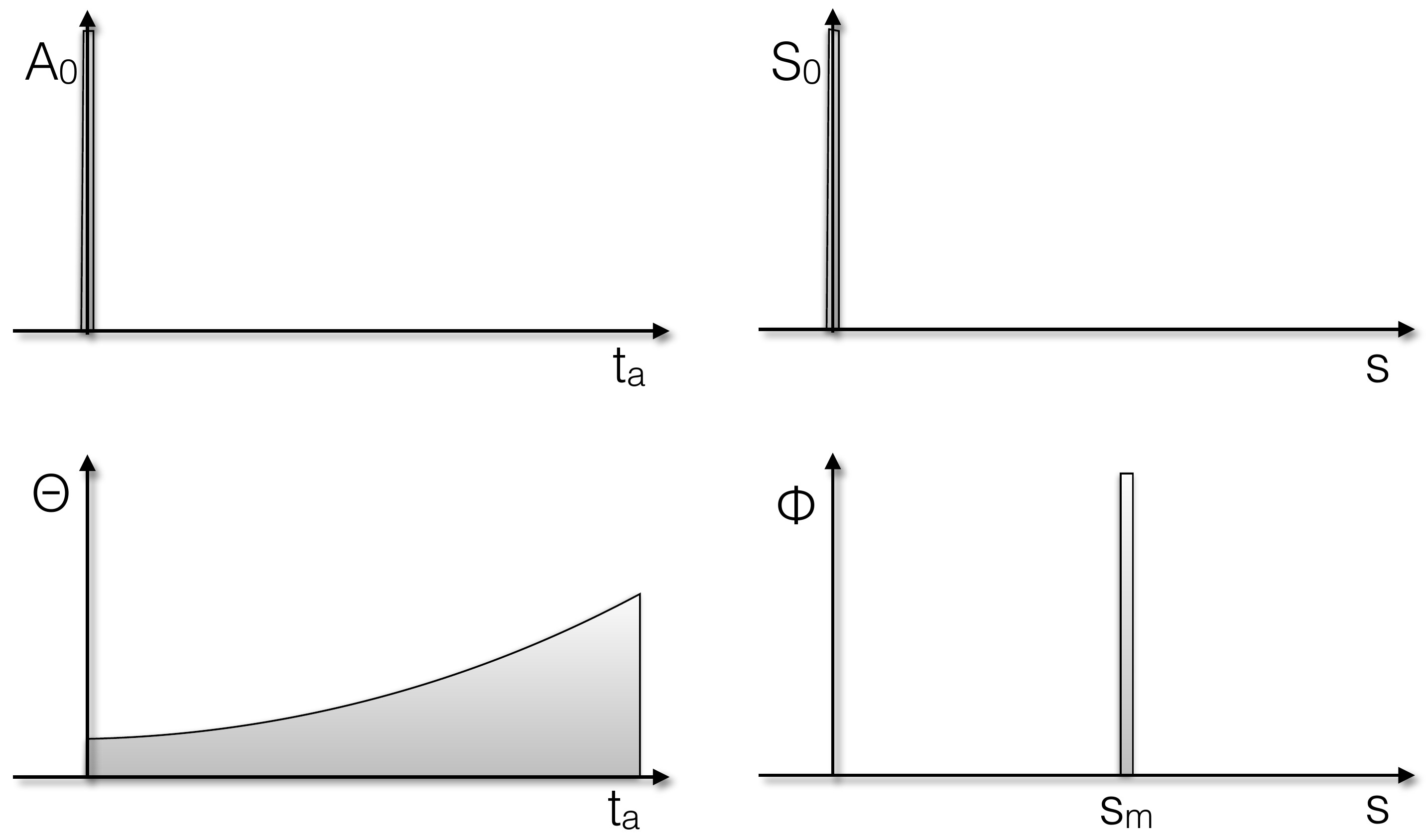}
\caption{We study a subset of the full model sketched in \figurename~\ref{fig:general_model_description} in which $A_0$, $S_0$ and $\Phi$ take their simplest forms:
$\Phi(s) = \delta (s-\smax)$, $A_0(\ta) = \delta (\ta-0)$ and $S_0(s) = \delta(s-0)$. This simplified model applies for all $\Theta$.
\label{fig:special_model_description}}
\end{figure}

To use the general framework we need to define the four underlying functions that are shown in \figurename~\ref{fig:general_model_description}. $\Theta(t_a)$ and $\Phi(s)$ govern the flow of probability out of $A$ and $S$, respectively. The other two, $A_0(t_a)$ and $S_0(s)$, control the flow of probability into $A$ and $S$, respectively. It is out of the scope of this paper to describe the effects of all four of them. Instead we focus on the effects of $\Theta$, which controls the transition from the slipping state to the pinned state, by reducing the other three functions to their simplest forms (\figurename~\ref{fig:special_model_description}).

Braun and Peyrard have studied the effects of the strength distribution (equivalent to $\Phi(s)$) extensively. We recommend their papers to interested readers and take as our first simplification here that $\Phi(s)$ is collapsed to a delta function at a maximum stretching threshold $\smax$,
\begin{align}
\Phi(s) = \delta(s-\smax).\label{eq:Phi_is_heaviside}
\end{align}
When all contacts have the same breaking threshold the portion of contacts that break during a time interval $\dt$ is reduced to
\begin{align}
P_\text{out of S} = \int_{\smax-\dx}^\smax S(s,t)\,\ds &\approx S(\smax,t)\,\dx.
\end{align}
We define $s$ to be positive for a positive displacement of the slider. Second, we collapse the distribution of initial slipping times, $A_0$, to a delta function at zero slipping time: $A_0(\ta) = \delta (\ta)$. Third, we collapse the distribution of initial stretchings, $S_0$, to a delta function at zero stretching: $S_0(s) = \delta(s)$.

This leaves only one remaining underlying function: $\Theta(t_a)$. The fraction of junctions that enter the pinned state during a time interval $\dt$ is
\begin{align}
P_\text{out of \A} = \dt \int_0^{\infty} \A(\ta,t) \Theta(\ta) \dta.
\end{align}
This probability should be removed from $A$.
Combining $P_\text{out of S}$, $P_\text{out of A}$ and the rigid shift in $S$ we write
\begin{align}
\begin{split}
&S(s,t+\dt) =\\
&
\left \{
  \begin{array}{l l}
    S(s-\dx,t)  + 
    \left \{
      \begin{array}{l l}
        \frac{P_\text{out of A}}{\dx} &, s \in [0, \dx]\\
	0 & , \text{otherwise}  \\
      \end{array}
    \right \}
      &, |s| \leq \smax \\
    0 &, |s| > \smax.\\
  \end{array}
\right.
\label{eq:S}
\end{split}
\end{align}
Note that for convenience we avoid $\delta$-functions in $S$ by distributing the junctions entering $S$ uniformly in the interval $s \in [0,\dx]$. Also note that the junctions can break at both $-\smax$ and $\smax$. The corresponding equation for $A$ is slightly different (the equations are formally equivalent in the full framework, but here we have specified $\Phi$ in equation~\eqref{eq:Phi_is_heaviside} while $\Theta$ remains unspecified).
\begin{align}
\begin{split}
&\A(\ta,t+\dt) =\\
&
\left \{
  \begin{array}{l l}
    \A(\ta-\dt,t)\left (1 - \dt \Theta(\ta) \right )&, \ta \in [\dt, \infty]\\
    \frac{P_\text{out of S}}{\dt} &, \ta \in [0, \dt]. \\
  \end{array}
\right.
\label{eq:A}
\end{split}
\end{align}
Also note that the velocity of the slider is
\begin{equation}
v(t) = \frac{\dx (t)}{\dt},
\end{equation}
which we will be useful in the next sections.

This concludes the description of the model we will use in the rest of the paper to study disorder in the time-dependent junction state. In sections \ref{sec:Steady state} and \ref{sec:transients} we will discuss the implications this time-disorder has at the macroscopic scale. We start with the steady state, where the slider moves at constant velocity.

\section{Steady state\label{sec:Steady state}}
In this section we find expressions for the distributions $S$ and $A$ during steady sliding and derive the steady state friction coefficient as a function of velocity.
The results are valid under the assumptions made on $\Phi$, $S_0$ and $A_0$ in the previous section. Additionally, we need to assume that a steady state can be found, which means that the block is sliding at constant velocity $v$ and that $\Phi$ and $\Theta$ are independent of the block's position and of global time $t$. 

\subsection{Steady state distributions}
The distance travelled by the block while a contact is in $S$ is
\begin{equation}
\text{distance travelled in S} = \smax,
\end{equation}
since we have already assumed that the distribution of initial stretchings $S_0$ is a delta function at zero stretching and that $\Phi$ breaks all contacts at $\smax$. The mean distance travelled while the contact is in $A$ is
\begin{equation}
\langle \text{distance travelled in A}\rangle = v\langle t_a\rangle,
\end{equation}
where $\langle t_a\rangle$ is the average time spent by a junction in the slipping state, i.e. the expected lifetime in the $A$ distribution. This is the expectation value of $t_a$ in the distribution of delay times (mapping from $\Theta$ in Appendix~\ref{appsec:Theta_equiv_tau}), and should not be confused with the expectation value of $t_a$ in $A$, $\int_0^\infty t_a A(t_a) dt_a$. The fraction of junctions in $S$, $P_S$, is then
\begin{align}
\begin{split}
P_S& = \frac{\text{distance travelled in S}}{\text{distance travelled in S}+\langle \text{distance travelled in A}\rangle} \\
&= \PS.
\end{split}
\end{align}
Similarly, the fraction of contacts in $A$, $P_A$, is
\begin{align}
\begin{split}
P_A& = \frac{\langle \text{distance travelled in A} \rangle}{\text{distance travelled in S}+\langle \text{distance travelled in A}\rangle} \\
&= \PA.\label{eq:PA}
\end{split}
\end{align}
In steady state, the exchange of probability between the distributions $A$ and $S$ is constant in time, otherwise the distributions would change. The same probability enters $S$ at $s=0$ and leaves $S$ at $s=\smax$:
\begin{equation}
P_\text{entering S} = P_\text{out of S}.
\end{equation}
With constant $v$ this is only possible when the shape of $S$ is uniform. Since $S$ is nonzero only on $s\in[0,\smax]$ we conclude that
\begin{align}
S_\text{steady state}(s) = 
  \underbracket{\PS}_{P_S}
  \underbracket{\left\{
    \begin{array}{ll}
      \frac{1}{\smax}	&, s\in[0,\smax]\\
      0			&, \text{else}.
  \end{array}\right.}_{\text{shape}}\label{eq:S_steady_state}
\end{align}

The calculation for $A$ with a general $\Theta$ is longer. It can be found in Appendix~\ref{appsec:AInSteadyState}, and we give the results for $\langle t_a\rangle$ and $A$ here.
\begin{equation}
\langle t_a \rangle = \int_0^\infty e^{-\int_0^{t_a} \Theta(t_a') \, \dt_a'} \,\dt_a,
\label{eq:expected_lifetime_in_A}
\end{equation}
\begin{equation}
A_\text{steady state} (t_a) = 
  \underbracket{\PA}_{P_A}
  \underbracket{\frac{e^{-\int_0^{t_a} \Theta(t_a') \,\dt_a'}}{\langle t_a \rangle}}_{\text{shape}}.  
\label{eq:A_steady_state}
\end{equation}
Note that the steady state distributions do not depend on $\nu_S$ or $\nu_A$.

\subsection{Steady state friction coefficient}
The steady state friction coefficient can be found using equation \eqref{eq:total_friction}. Inserting for $A$ and $S$ we find
\begin{align}
\begin{split}
& \mu_\text{steady state} (v) \equiv \nu_\text{macro,steady state} (v) =  \\
& \int_{0}^{\smax} \frac{ \nu_S(s)}{\smax+v\langle t_a \rangle} \,\ds + \int_0^\infty  \frac{\nu_A(\ta) v e^{-\int_0^{t_a} \Theta(t_a') \,\dt_a'}}{\smax+v\langle t_a \rangle} \,\dta.
\end{split}
\label{eq:total_friction_steady_state}
\end{align}
Equation \eqref{eq:total_friction_steady_state} holds for any choice of $\Theta$, $\nu_S$ and $\nu_A$, which results in a large variety of possible steady state friction laws. To find the steady state friction law for a particular system, one must specify $\Theta$, $\nu_S$ and $\nu_A$ based on the physical properties of the individual junctions. We will give a couple of examples of such single junction behavior laws and show that monotonous as well as non-monotonous velocity-dependent steady-state friction laws can be found within the model. We consider successively the cases of a velocity-independent and a velocity-dependent $\Theta$.

\subsubsection{No velocity dependence in \texorpdfstring{$\nu_S$, $\nu_A$ or $\Theta$}{nu\_S, nu\_A or Theta}}
\begin{figure}
\centering
\includegraphics[width = .49\textwidth]{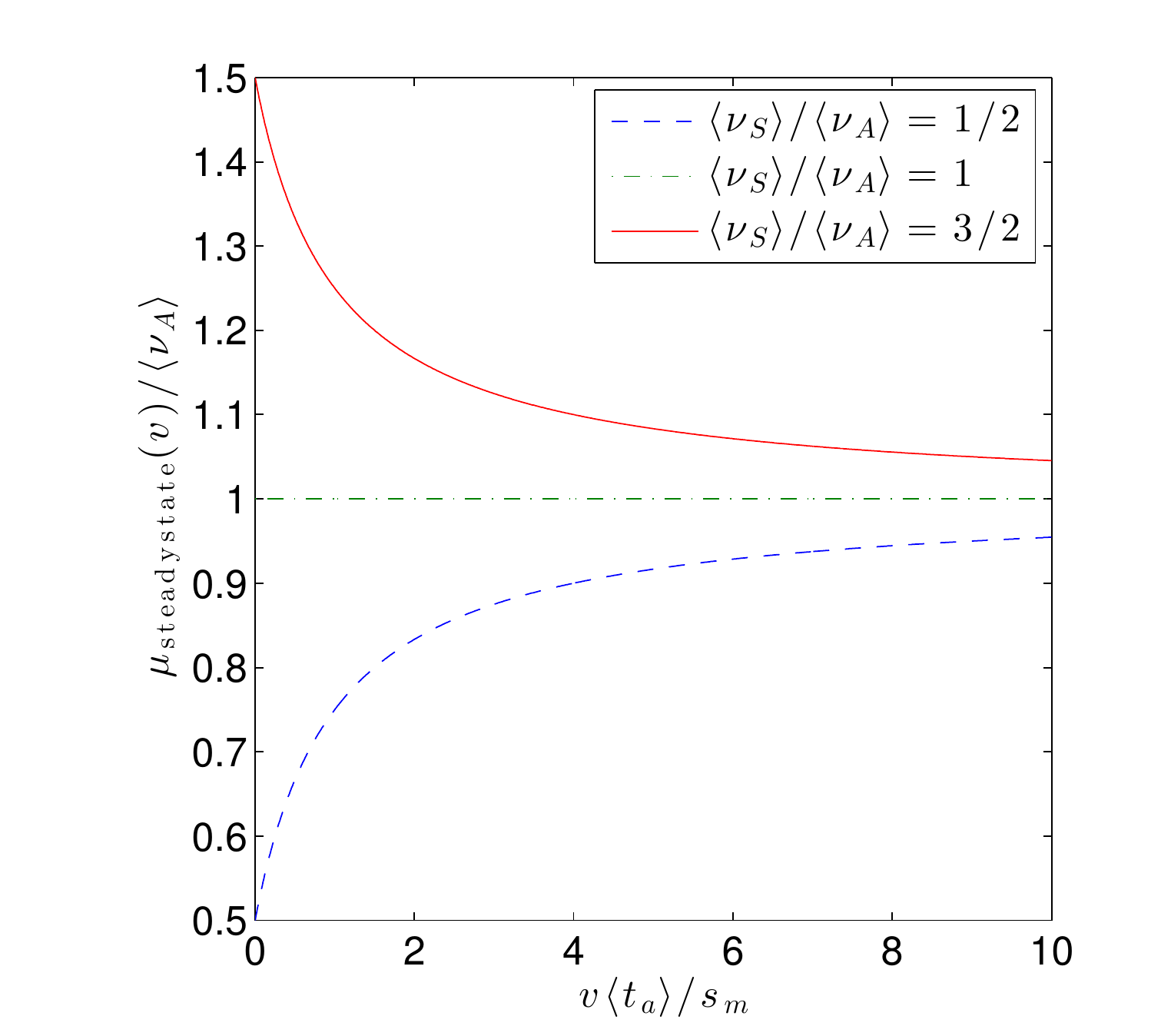}
\caption{(Color online) Velocity dependence of the steady state friction coefficient as given in equation \eqref{eq:steady_state_friction_coeff_velocity_independent_theta}. When $\nu_S$, $\nu_A$ and $\Theta$ are all velocity independent, $\mu_\text{steady state}(v)$ is monotonic. It is velocity weakening if $\langle \nu_S \rangle > \langle \nu_A \rangle$, velocity strengthening if $\langle \nu_S \rangle < \langle \nu_A \rangle$, and constant if $\langle \nu_S \rangle = \langle \nu_A \rangle$. In all cases it converges to $\langle \nu_A \rangle$ for $v \rightarrow \infty$.
\label{fig:steady_state_friction_coefficient_velindependent}}
\end{figure}

When $\Theta$ is velocity-independent, the shapes of $A$ and $S$ are also velocity-independent, and so is $\langle t_a \rangle$. In these cases the velocity only controls the amount of probability in each of the distributions, i.e. the amplitudes $P_S$ and $P_A$. The shape of $A$, however, can still be different for different $\Theta$. We define the following velocity-independent expectation values:
\begin{align}
\langle \nu_S\rangle &= \int_0^\smax \frac{\nu_S(s)}{\smax}\,\ds,\\
\langle \nu_A\rangle &= \int_0^\infty\frac{\nu_A(\ta)e^{-\int_0^\ta\Theta(\ta')\,\dta'}}{\langle t_a\rangle}\,\dta.
\end{align}
It follows from equation~\eqref{eq:total_friction_steady_state} (or alternatively from equations~\eqref{eq:total_friction}, \eqref{eq:S_steady_state} and \eqref{eq:A_steady_state}) that 
\begin{align}
\mu_\text{steady state} (v) 
& = P_\text{S} \langle \nu_S \rangle + P_\text{A} \langle \nu_A \rangle \label{eq:mu_steady_state_splitting}\\
& = \PS \langle \nu_S \rangle + \PA \langle \nu_A \rangle.
\end{align}
We can write this in terms of the dimensionless parameters $\frac{\mu_\text{steady state} (v)}{\langle \nu_A \rangle}$, $\frac{v \langle t_a \rangle}{\smax}$ and $\frac{\langle \nu_S \rangle}{\langle \nu_A \rangle }$:
\begin{align}
\frac{\mu_\text{steady state} (v)}{\langle \nu_A \rangle} = \frac{1}{1 + \frac{v \langle t_a \rangle}{\smax} }    \left ( \frac{\langle \nu_S \rangle}{\langle \nu_A \rangle } + \frac{v \langle t_a \rangle}{\smax} \right ).
\label{eq:steady_state_friction_coeff_velocity_independent_theta}
\end{align}
The velocity weakening, velocity strengthening and velocity independent solutions are apparent from this form. Increasing the velocity shifts probability from $S$ to $A$, which results in a velocity weakening steady state friction coefficient if $\langle \nu_S \rangle > \langle \nu_A \rangle$, a velocity strengthening steady state friction coefficient if $\langle \nu_S \rangle < \langle \nu_A \rangle$, and a constant steady state friction coefficient if $\langle \nu_S \rangle = \langle \nu_A \rangle$. For all three cases, the $v \rightarrow \infty$ limit is $\langle \nu_A \rangle$. This is demonstrated in \figurename~\ref{fig:steady_state_friction_coefficient_velindependent}.

\subsubsection{Velocity-dependent \texorpdfstring{$\Theta$}{Theta}}
If $\Theta$ is velocity-dependent the shape of $A$ will also be velocity-dependent, and equation~\eqref{eq:total_friction_steady_state} can no longer be factorized as in the previous section. There are many possible choices for $\Theta$. Here we give an example motivated from a realistic microscopic picture of micro-contacts at the interface between rough solids, which results in a weakening-then-strengthening velocity-dependent steady state friction law.
\begin{figure}
\centering
\includegraphics[width = .35\textwidth]{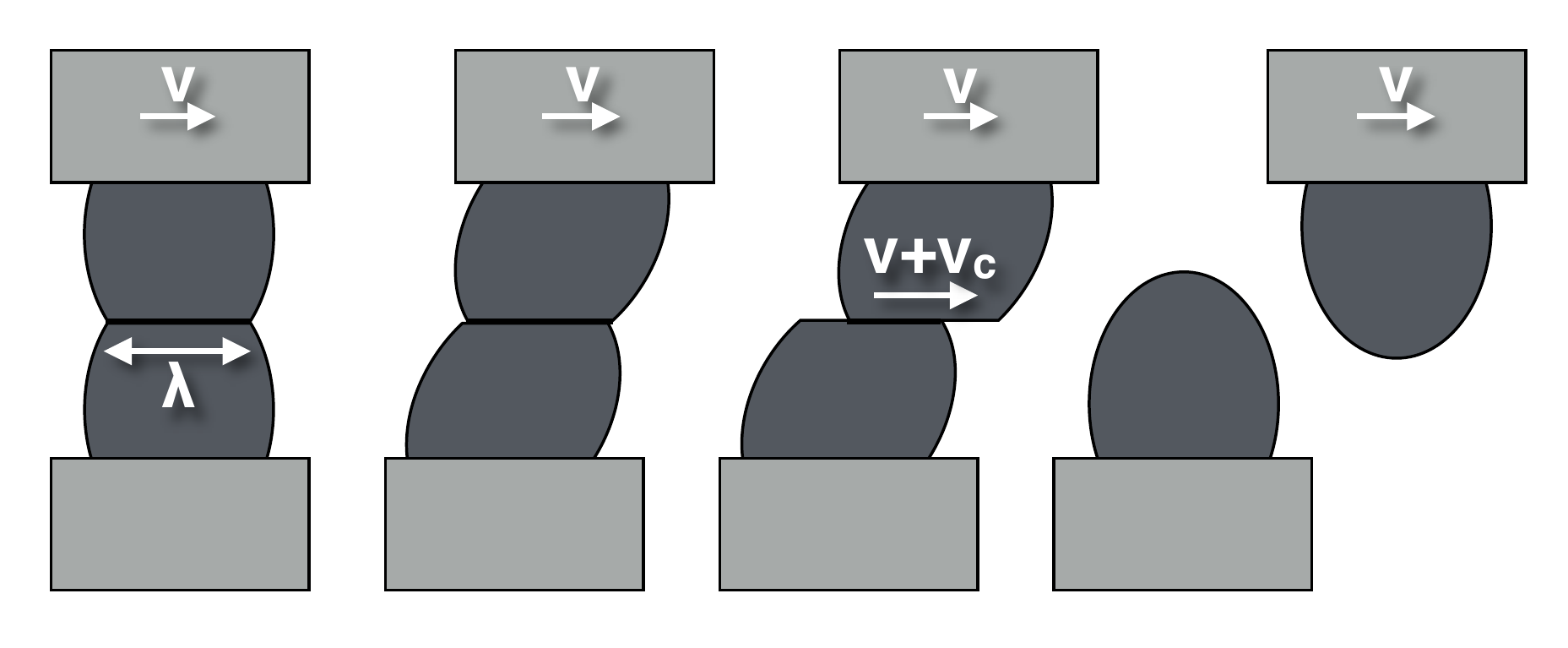}
\caption{
Sketch of a realistic micro-junction behavior. \emph{Left}: the micro-junction is a micro-contact of size $\lambda$ between two asperities. \emph{Middle left}: the pinned state is linear elastic, with $\nu_S(s) = ks$. When the stretching reaches its threshold, $\smax$, the junction enters a slipping state. \emph{Middle right}: it slides a characteristic distance $\lambda$ before it vanishes (\emph{right}): it leaves the slipping state and is replaced by a new junction (not shown). The net slipping velocity at the micro-contact interface is the sum of a characteristic creep-like velocity $v_c$ and the slider velocity $v$.\label{fig:velocity_dep_friction_sketch}}
\end{figure}

Assume the following micro-contact behavior: The pinned state is linear elastic so that $\nu_S(s) = ks$, where $k$ is an elastic constant with dimension 1/length. When a contact reaches the threshold $\smax$, it enters the slipping state, in which it will slide a characteristic distance $\lambda$ before the micro-contact vanishes (see \figurename~\ref{fig:velocity_dep_friction_sketch}). We assume that, even in the absence of driving velocity $v$, the contact, once in the slipping state, would slip with a small velocity $v_c$, as observed experimentally in e.g. \cite{Heslot1994creep, Yang2008dynamics} and attributed to a creep-like process. The net velocity at the micro-contact interface in the slipping state is the sum of $v_c$ and of the macroscopic block velocity $v$, giving
\begin{align}
\text{distance slipped} = v_c \Delta t + \int_0^{\Delta t} v(t) \,\dt.
\end{align}
This is illustrated in \figurename~\ref{fig:velocity_dep_friction_sketch}. We seek the time interval $\Delta t$ where the distance slipped equals the characteristic distance $\lambda$. Note that due to the creep $v_c$, this occurs even if the slider velocity is zero (after a characteristic time $t_c = \frac{\lambda}{v_c}$). In steady state, the velocity is constant, and we can solve for
\begin{align}
\Delta t = \frac{\lambda}{v_c + v}.
\end{align}
To relate $\Delta t$ to $\Theta$ we use a $\Theta$ with
\begin{equation}
\Delta t = \langle t_a \rangle = \int_0^\infty e^{-\int_0^\ta \Theta(t_a',v) \,\dta'} \,\dta.
\end{equation}
Of the many choices of $\Theta$ that would work here, we choose the simplest one and assume that $\Theta$ is not a function of $t_a$. We get
\begin{equation}
\frac{\lambda}{v_c + v}= \int_0^\infty e^{-\ta \Theta(v)} \,\dta = \frac{1}{\Theta(v)},
\end{equation}
which gives us
\begin{equation}
\Theta(v) = \frac{v_c + v}{\lambda}. \label{eq:theta_v}
\end{equation}
Inserting $\Theta$ in equation \eqref{eq:expected_lifetime_in_A} gives
\begin{align}
\langle t_a \rangle = \frac{\lambda}{v_c + v}.
\end{align}

\begin{figure}
\centering
\includegraphics[width = .49\textwidth]{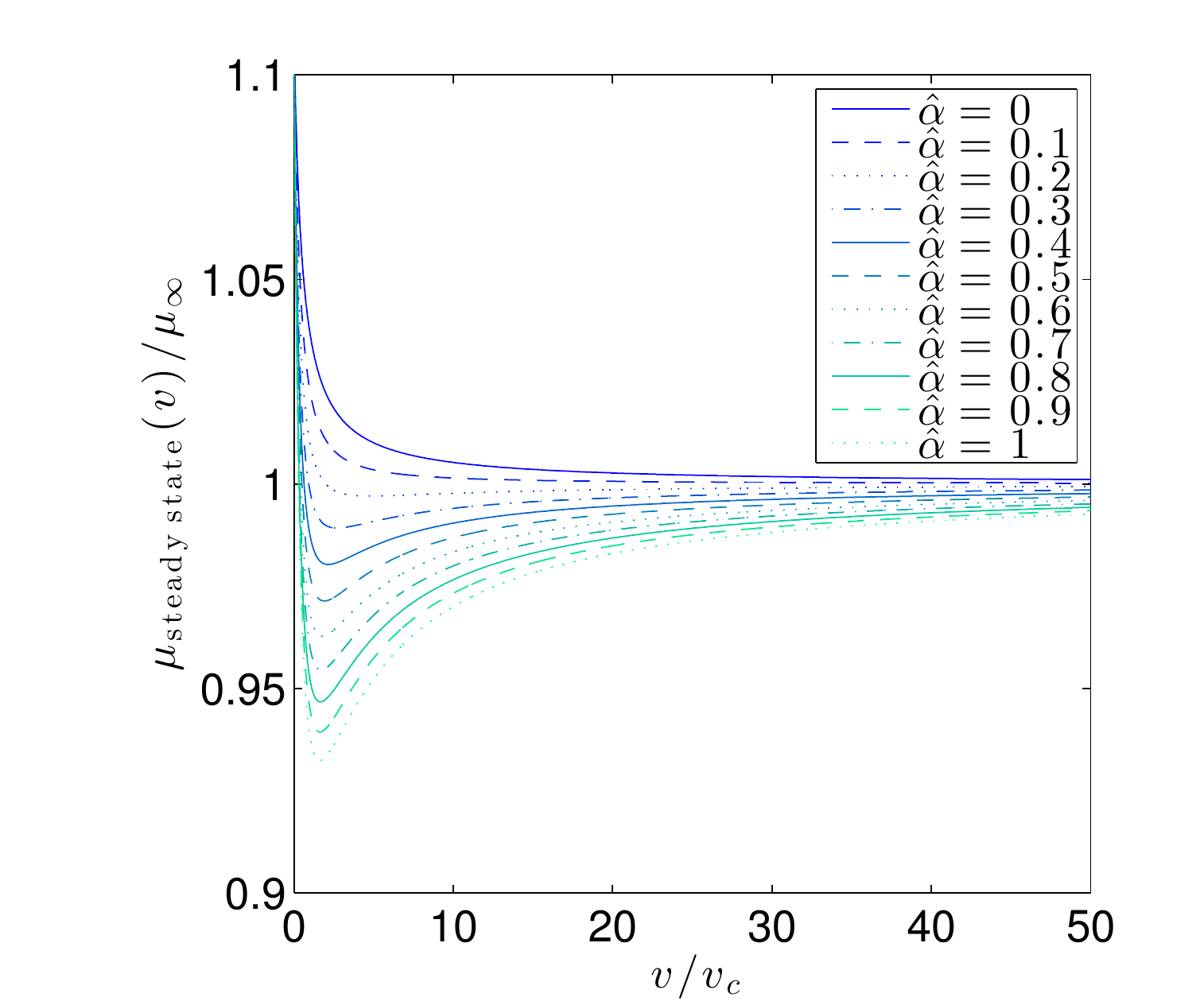}
\caption{
(Color online) Velocity dependence of the steady state friction coefficient as given in equation \eqref{eq:steady_state_friction_coefficient_velocity_dependent_theta_dimensionless}. When $\Theta$ is velocity-dependent (equation \eqref{eq:theta_v}) and $\nu_A$ is time-weakening (equation \eqref{eq:nua}), the steady state friction coefficient has solutions where $\mu_\text{steady state}$ is initially weakening, then strengthening and converging to a constant level $\mu_\infty$ (equation \eqref{eq:mu_infty}) when $v \rightarrow \infty$. The figure shows the particular case $\nu_0 = 0.4k\smax$, for various $\hat \alpha = \alpha \lambda/v_c$, increasing from blue to green (dark to light gray).
\label{fig:steady_state_friction_coefficient_veldependent}}
\end{figure}

To complete the friction law we need $\nu_A$. Assume that the force from each micro-contact decreases with the time spent in the slipping state, for example due to friction induced heating. We take as our example
\begin{equation}
\nu_A(t_a) = \nu_0 e^{-\alpha t_a},\label{eq:nua}
\end{equation}
where $\alpha$ is a positive constant with dimension 1/time. 
We can determine the steady state friction coefficient by inserting $\nu_S$, $\nu_A$ and $\Theta$ into equation \eqref{eq:total_friction_steady_state}. We get
\begin{align}
\begin{split}
&\mu_\text{steady state} (v) = \\
& \frac{1}{\smax+\frac{v \lambda}{v_c + v}}  \left (  \frac{1}{2} k\smax^2 +      \frac{\nu_0}{  \frac{v_c + v}{v\lambda} + \alpha/v}      \right ).
\end{split}
\end{align}
This can be written in terms of the dimensionless quantities $\hat s = \smax/\lambda$, $\hat k = k\lambda$, $\hat \alpha = \alpha\lambda/v_c$ and $\hat v = v/v_c$: 
\begin{align}
\mu_\text{steady state} (\hat v) =  \frac{1}{\hat s+\frac{\hat v}{1 + \hat v}}  \left (  \frac{1}{2} \hat k \hat s^2 +      \frac{\nu_0 \hat v}{ 1 + \hat \alpha +  \hat v}      \right ).
\label{eq:steady_state_friction_coefficient_velocity_dependent_theta_dimensionless}
\end{align}
Equation \eqref{eq:steady_state_friction_coefficient_velocity_dependent_theta_dimensionless} is plotted in \figurename~\ref{fig:steady_state_friction_coefficient_veldependent} for different values of $\alpha$. Note that $\hat k \hat s = k \smax$, which is directly related to the force threshold in the pinned state. In a physical system, the force in the slipping state is usually smaller than or equal to the threshold in the pinned state: $k \smax \geq \nu_0$. Depending on the value of $\alpha$, $\mu_\text{steady state}(v)$ can be both velocity-weakening at low velocities and velocity-strengthening at high velocities, with a transition velocity $\hat v_\text{transition}$ given in appendix \ref{app:steady_state_vel_dep_theta}. When $v \rightarrow \infty$, $\mu_\text{steady state}$ converges towards
\begin{align}
\mu_\infty = \lim_{\hat v\rightarrow \infty} \mu_\text{steady state} (\hat v) = \frac{\frac{1}{2} \hat k \hat s ^2 + \nu_0}{\hat s + 1},
\label{eq:mu_infty}
\end{align}
which is different from $\langle \nu_A \rangle$. For the velocity-independent $\Theta$ of the previous section, all junctions were $A$ at $v\rightarrow \infty$ ($P_A = 1$ and $P_S = 0$). In the present case, however, $P_S$ and $P_A$ converge to values different from these because $\langle t_a \rangle \rightarrow 0$ when $v \rightarrow \infty$.
\begin{align}
\lim_{v \rightarrow \infty} P_S = \frac{\smax}{\smax + \lambda}
\end{align}
and
\begin{align}
\lim_{v \rightarrow \infty} P_A  = \frac{\lambda}{\smax + \lambda}.
\end{align}

Weakening-then-strenthening velocity-dependent friction laws are often considered to be generic for frictional interfaces \cite{Baumberger2006solid, bar2013velocity}. It was observed experimentally on interfaces as different as paper-on-paper \cite{Heslot1994creep}, granite-on-granite \cite{Dieterich1978Time} or PMMA-on-glass \cite{Bureau2002rheological}. It was used as a mesoscopic friction law in numerical models of the statistics of earthquakes at seismic faults \cite{burridge1967model}. It was also discussed theoretically in \cite{BarSinai2012slow, bar2013velocity, BarSinai2013instabilities}. Note that Braun and Peyrard found an opposite behaviour, strengthening-then-weakening, when using ageing  and delay times \cite{Braun2011dependence}.

\section{History dependent friction \label{sec:transients}}
When the velocity is not constant, the slider will in general not be in the steady state. 
While steady state solutions are reasonable approximations in some limits, an understanding of transients is particularly important to the study of a large variety of frictional situations including oscillating contacts \cite{Rigaud2013interfacial}, the onset of frictional sliding, be it quasi-static \cite{Prevost2013probing} or dynamic \cite{Rubinstein2004detachment}, the cessation of slip \cite{Ben-David2010slip-stick} and friction instabilities \cite{LeRouzic2013squeal}.

In this section we will demonstrate two important consequences of a two-state junction law: slip history dependent static friction and slow slip. 
First, due to the coupling between the time-dependent state and the displacement dependent state, the distribution of junction stretchings, $S$, depends on the velocity history of the slider. In turn, this distribution determines the static friction coefficient during onset of sliding, resulting in history dependent static friction. In particular, the static friction coefficient depends on the deceleration when the slider last came to rest.  Second, if there is a weakening in the friction force in the slipping state that is distributed over a characteristic time (e.g. $\langle t_a \rangle$), and if $\nu_A \neq 0$, then slow slip is predicted from the model. 

To arrive at these promised results, we first need to develop a framework to make analytical predictions for transients. This is done in section \ref{sec:theoretical_bg}.  The applications are discussed in sections \ref{sec:history_dependent_static_friction} and \ref{sec:slow_slip}, which contain the main results of this paper.

\subsection{Analytical predictions for transients \label{sec:theoretical_bg}}
In this section we introduce the necessary framework to make analytical predictions of the friction force during transients. We start with  the distribution of junction stretchings, $S$, after stopping. We then relate $S$ to the mesoscopic static friction coefficient. The distribution of junctions in the slipping state, $A$, after onset of sliding can be found in Appendix \ref{sec:A_after_onset_of_slip}.

\subsubsection{Calculating \texorpdfstring{$S(s,t)$}{S(s,t)} after stopping}
As we will see in Section~\ref{sec:MesoscopicStaticFrictionCoefficient}, $S(s,t)$ is of particular interest because it determines the development of the friction coefficient when sliding is initiated. In this section we will find $S(s,t)$ after motion stops. Our strategy is to find $P_\text{out of A}$ as a function of time, and place it in $S$ according to the velocity profile. 

Given an initial state $A(t_a,0)$, the probability in $A$ follows
\begin{equation}
P_A(t) = \int_0^\infty A(t_a,0) e^{-\int_{t_a}^{t+t_a} \Theta(t_a') \,\dt_a'}\, \dt_a
\end{equation}
where $e^{-\int_{t_a}^{t+t_a} \Theta(t_a') \,\dt_a'}$ is the fraction of junctions, that had been slipping for a time $t_a$ at $t=0$, that remain in $A$ after a time $t$. This is found from solving equation \eqref{eq:PAspike_diff_equation} for the evolution of junctions in the slipping state.
The cumulative probability that has left $A$ is 
\begin{align}
&P_A(0)-P_A(t) = \notag\\
&\int_0^\infty A(t_a,0) \,\dt_a - \int_0^\infty A(t_a,0) e^{-\int_{t_a}^{t+t_a} \Theta(t_a') \,\dt_a'} \,\dt_a.
\label{eq:cum_prob_from_A}
\end{align}
The instantaneous probability leaving $A$ is found by taking the derivative of equation \eqref{eq:cum_prob_from_A} with respect to $t$ and multiplying this with $\dt$.
\begin{equation}
P_\text{out of A}(t)= \frac{\text{d}}{\dt} \left (- \int_0^\infty A(t_a,0) e^{-\int_{t_a}^{t+t_a} \Theta(t_a') \,\dt_a'} \,\dt_a \right ) \,\dt.
\end{equation}
Dividing by $\dt$ we get
\begin{align}
\begin{split}
&P_\text{out of A}(t)/\dt \\
& =   \int_0^\infty A(t_a,0)  \frac{\text{d}}{\dt} \left ( - e^{-\int_{t_a}^{t+t_a} \Theta(t_a') \,\dt_a'}\right ) \,\dt_a \\
& =   \int_0^\infty A(t_a,0)   \left (\frac{\text{d}}{\dt}  \int_{t_a}^{t+t_a} \Theta(t_a') \,\dt_a' \right ) e^{-\int_{t_a}^{t+t_a} \Theta(t_a') \,\dt_a'} \,\dt_a.
\end{split}
\end{align}
Defining the rate of probability from $A$ to $S$, $R_{A\rightarrow S}(t) \equiv P_\text{out of A}(t)/\dt $, and using Leibniz' integration rule on $\frac{\text{d}}{\dt} \int_{t_a}^{t+t_a} \Theta(t_a') \,\dt_a'$ we find
\begin{align}
R_{A\rightarrow S}(t) =   \int_0^\infty A(t_a,0)  \Theta(t+t_a) e^{-\int_{t_a}^{t+t_a} \Theta(t_a') \,\dt_a'} \,\dt_a. 
\label{eq:P_OutOfA_AsAFunctionOfTime}
\end{align}
We can find the stretching distribution after the slider has stopped (at time $t_\text{stop}$), $S(s,t_\text{stop})$, if we assume that no junctions reach $\smax$ and break during the time $t_\text{stop}$, i.e. when the total displacement $\Delta x(t_\text{stop})$ is smaller than $\smax$. Recall from equation~\eqref{eq:S} that probability enters $S$ with amplitude $P_\text{out of A}/\dx = R_{A\rightarrow S}/v$ (which results in a velocity dependence in $S$ after stopping) and combine this with the stretching that occurs when the probability enters $S$ and until $t_\text{stop}$ to obtain
\begin{align}
\begin{split}
& S(\Delta x(t_\text{stop})-\Delta x (t),t_\text{stop})= \\
& \frac{1}{v(t)}  \int_0^\infty A(t_a,0) \Theta(t+t_a) e^{-\int_{t_a}^{t+t_a} \Theta(t_a') \,\dt_a'} \,\dt_a, 
\end{split}
\label{eq:S_AfterStoppingAnalytical}
\end{align}
where $\Delta x(t)$ is the displacement of the slider. 
Note that there is a time-dependence here, while $S$ is in general given as a function of stretching. To go from $t$ to $s$, we need the velocity as the slider comes to rest. We show an example of such mapping in section \ref{sec:history_dependent_static_friction}. Equation \eqref{eq:S_AfterStoppingAnalytical} applies for any initial state where $P_A = 1$, and for any $\Theta$. Note that $S$ depends on the velocity the slider has as it comes to rest through the term  $\frac{1}{v(t)}$, which has implications for the static friction coefficient assosiated the next onset of sliding.

\subsubsection{Macroscopic static friction coefficient\label{sec:MesoscopicStaticFrictionCoefficient}}
In this section we show how the macroscopic static friction coefficient depends on the distribution of forces acting on the junctions, $S(s)$.
Consider a slider at rest where $S(s)$ is known. At any instant in time, the friction force is
\begin{equation}
\nu = \int_{0}^{\smax} \nu_S(s) S(s) \ds + \int_0^\infty \nu_A(t_a)A(t_a) \,\dt_a.
\end{equation}
If we assume that breaking is fast compared to $\langle t_a \rangle$, the friction force as a function of displacement $\Delta x$ is
\begin{align}
\begin{split}
\nu(\Delta x) \approx & \int_{0}^{\smax-\Delta x} \nu_S(s+\Delta x) S(s) \,\ds\\
& + \nu_A(0)\int_{\smax-\Delta x}^{\smax} S(s) \,\ds,
\end{split}
\label{eq:mu_x_fiber_bundle}
\end{align}
where the integral in the second term is the probability in $A$, $P_A$. $\nu_A(0)\int_{\smax-\Delta x}^{\smax} S(s) \,\ds$ should be replaced with an integral over $A$ from equation \eqref{eq:A_initial_after_slip} or be solved numerically if the assumption that breaking is fast compared to $\langle t_a \rangle$ does not hold. This would require knowledge of the velocity of the slider during breaking. In the following we assume that equation \eqref{eq:mu_x_fiber_bundle} is a good approximation.
The macroscopic static friction coefficient is simply the maximum of $\nu(\Delta x)$:
\begin{align}
\begin{split}
\mu_s & =  \text{max} (\nu(\Delta x))\\
= &\text{max}  \left( \int_{0}^{\smax-\Delta x} \nu_S(s+\Delta x) S(s) \,\ds \right.\\
  &+           \left. \nu_A(0)\int_{\smax-\Delta x}^{\smax} S(s) \,\ds         \right).
\end{split}
\label{eq:fiber_bundle_friction_coefficient}
\end{align}
This concludes the discussion of the framework for analytical predictions of transients. We will now use these results in a few examples.

\subsection{Deceleration dependent macroscopic static friction \label{sec:history_dependent_static_friction}}

Equation \eqref{eq:fiber_bundle_friction_coefficient} predicts that the macroscopic static friction coefficient depends directly on $S$. The highest possible static friction is found when $S$ is a delta-function so that all junctions contribute their maximum force simultaneously,
\begin{align}
\mu_{s,\text{max}} = \nu_S(\smax).
\end{align}
Increasing the width of $S$ will reduce $\mu_s$ because the first junctions break before the rest reach their maximum force contribution.  When $\nu_A = 0$ and $\nu_S(s)$ is strictly increasing, the lowest static friction coefficient is obtained for $S(s)$ uniform on $s\in[0,\smax]$;
\begin{align}
\mu_{s,\text{min}} = \int_0^\smax \frac{\nu_S(s)}{\smax} \ds.\label{eq:mu_s_min}
\end{align}
When $\nu_S(s) = ks$, the ratio of maximum to minimum static friction (from the two limits of $S$) is $\frac{\mu_{s,\text{max}}}{\mu_{s,\text{min}}} = 2$. It is noteworthy that the value can vary this much even though all micro-junctions have the same breaking threshold. A similar result was already found by Farkas et al. \cite{Farkas2005static}. In our previous work \cite{Tromborg2014slow} we found the dependence of $\mu_s$ on the width of $S$ numerically. In appendix~\ref{appsec:staticFrictionFromUniformS} we find the same result analytically for uniform $S$ by solving equation~\eqref{eq:fiber_bundle_friction_coefficient}. Increasing the width of $S$ will in general reduce the static friction coefficient. However, the precise functional form of this dependence also depends on the shape of $S$.

Combining equation \eqref{eq:S_AfterStoppingAnalytical} and equation \eqref{eq:fiber_bundle_friction_coefficient}, we see that the macroscopic static friction coefficient depends on the velocity history, because $v(t)$ determines $S$, which in turn determines $\mu_s$. Even though the assumptions made when deriving equation \eqref{eq:S_AfterStoppingAnalytical} were quite restrictive, the dependence of friction on velocity history is general in our model. This velocity history dependence has interesting consequences: The macroscopic static friction coefficient depends on slip dynamics of the previous sliding event.

In this section we calculate $S$ for a slider that stops under constant deceleration. We then use this result to find the static friction coefficient of the next event as a function of the deceleration. Because the analytical results can only be found under quite strict assumptions, we will complement them with numerical solutions in which the restricting assumptions can be lifted. The implementation is straightforward using equation \eqref{eq:S} and \eqref{eq:A}.

Assume that all probability is initially in $A$; $P_A(0)=1$. As initial condition, $A(t_a,0)$, for the analytical calculations we choose the steady state distribution of $A$ when $v\rightarrow \infty$, which is a good approximation to the steady state result for high velocities. Also assume that motion stops within the displacement $\smax$. The calculations would be simplest for $\Theta=\text{constant}$. However, $\Theta = bt_a$ results in $S$ distributions after stopping that are easier to interpret, and so we use this $\Theta$ for clarity of presentation. Numerically we solve for three different $\Theta$ including a constant $\Theta$.

Inserting $\Theta=b\ta$ in equation~\eqref{eq:S_AfterStoppingAnalytical} we get
\begin{align}
\begin{split}
&S(\Delta x(t_\text{stop})-\Delta x (t),t_\text{stop})= \\
&\frac{1}{v(t)}  \int_0^\infty A(t_a,0) b(t + t_a) e^{- \frac{1}{2}b \left ( (t+t_a)^2 - t_a^2 \right )} \,\dt_a. 
\end{split}
\label{eq:S_after_stopping_theta_bt}
\end{align}
If we let $v\rightarrow \infty$ in equation~\eqref{eq:A_steady_state} we find
\begin{equation}
\begin{split}
A_{\text{steady state},\infty} (t_a) &=  \frac{e^{-\int_0^{t_a} \Theta(t_a') \,\dt_a'}}{\langle t_a \rangle} \\
&= \frac{e^{-\frac{1}{2}bt_a^2}}{\langle t_a \rangle},
\end{split}
\end{equation}
where,
\begin{equation}
\langle t_a \rangle = \int_0^\infty e^{- \frac{1}{2}bt_a^2} \,\dt_a = \sqrt{\frac{\pi}{2b}}
\end{equation}
from equation~\eqref{eq:expected_lifetime_in_A}. Inserting this in equation \eqref{eq:S_after_stopping_theta_bt} yields
\begin{align}
\begin{split}
&S(\Delta x(t_\text{stop})-\Delta x (t),t_\text{stop})\\
& = \frac{1}{v(t)}  \int_0^\infty \frac{e^{-\frac{1}{2}bt_a^2}}{\sqrt{\frac{\pi}{2b}}}  b(t + t_a) e^{- \frac{1}{2}b \left ( (t+t_a)^2 - t_a^2 \right )} \,\dt_a \\
& = \sqrt{\frac{2b}{\pi}} \frac{e^{-\frac{1}{2}bt^2} }{v(t)}.\label{eq:S(t,t_stop)}
\end{split}
\end{align}
We have found $S$, but the independent variable in the expression is time. To find $S$ as a function of $s$ we need to find $t(\Delta x)$ and the correspondence between $\Delta x$ and $s$. Numerically the inversion is trivial ($\Delta x$ and $t$ come in indexed pairs and either is known as a function of the other). The analytical inversion requires a bit of bookkeeping. Assume that the slider stops under a constant deceleration, $a<0$, so that
\begin{align}
v(t) = 
\left \{
  \begin{array}{l l}
    	v_0 + at, & t < -\frac{v_0}{a} \\
	0, &  t \geq -\frac{v_0}{a}.
	\end{array}
\right.
\end{align}
Then
\begin{align}
\Delta x = 
\left \{
  \begin{array}{l l}
    	v_0t + \frac{1}{2}at^2, & t < -\frac{v_0}{a} \\
	\frac{1}{2}\frac{v_0^2}{-a}, &  t \geq -\frac{v_0}{a},
	\end{array}
\right.
\end{align}
and
\begin{align}
t(\Delta x) = 
\left \{
  \begin{array}{l l}
  	\frac{v_0 - \sqrt{v_0^2 + 2a \Delta x}}{-a}, &\Delta x <  -\frac{1}{2}\frac{v_0^2}{a}\\
	\text{not well defined}, & \Delta x \geq -\frac{1}{2}\frac{v_0^2}{a}.
  \end{array}
\right.
\end{align}
Because the block stops at $t_\text{stop} = -\frac{v_0}{a}$ and then remains at $\Delta x_\text{stop} = \Delta x (t_\text{stop}) = -\frac{1}{2}\frac{v_0^2}{a}$ the inversion is not well defined for larger values of $\Delta x$. This is handled by realizing that all the probability that leaves $A$ after the block has come to rest will enter $S$ at $s=0$ because $S_0=\delta(s)$. In the end we will therefore use
\begin{align}
S(s,t>t_\text{stop}) = S(s,t_\text{stop})  + C \delta(s),\label{eq:S_after_stopping}
\end{align}
where $C$ is the probability that shifts from $A$ to $S$ in the time interval $[t_\text{stop},t]$,
\begin{align}
C = \int_{t_\text{stop}}^t P_\text{out of A} (t') \,\dt'.
\end{align}
For $t \gg t_\text{stop}$, when all the probability is in $S$,
\begin{align}
C = 1 - \int_0^{\Delta x(t_\text{stop})} S(s,t_\text{stop}) \,\ds.
\end{align}
$S(s)$ is found by inserting
\begin{align}
&v(t(\Delta x)) = \sqrt{v_0^2 + 2a \Delta x}
\end{align}
and
\begin{align}
&s = \Delta x(t_\text{stop})-\Delta x  \Rightarrow \Delta x  =  -\frac{1}{2}\frac{v_0^2}{a} - s
\end{align}
in equation~\eqref{eq:S(t,t_stop)}. We obtain 
 \begin{align}
 S(s,t_\text{stop}) = 
 \left \{
   \begin{array}{l l}
 	\sqrt{-\frac{b}{\pi as}} e^{-\frac{b}{2a^2}(-v_0 + \sqrt{-2as})^2} &, s \in [0,\Delta x_\text{stop}] \\
 	0 &, \text{else}.
   \end{array}
 \right.    
 \label{eq:S_after_stopping_theta_bt_function_of_s}
 \end{align}

\begin{figure}
\centering
\includegraphics[width=.49\textwidth]{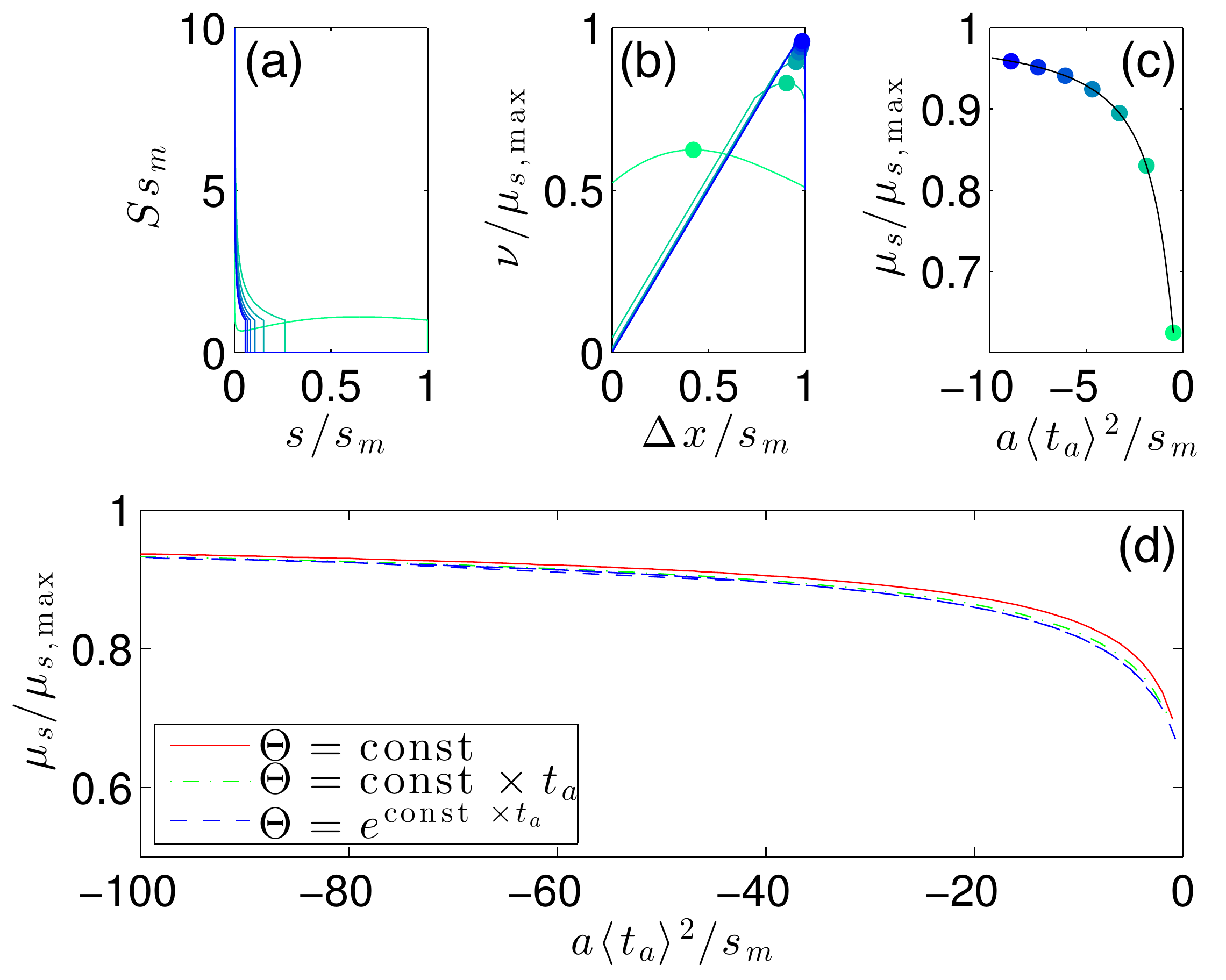}
\caption{(Color online) Macroscopic static friction coefficient depends on the stopping dynamics of the previous slip event. Equal colors in (a)-(c) correspond to the same parameter values (deceleration increasing from green to blue (light to dark gray)). (a) Junction stretching distribution $S(s)$ after slip cessation with constant deceleration (equation \eqref{eq:S_after_stopping_theta_bt_function_of_s}, where we assumed $\Theta = bt_a$). The initial condition is $P_A=1$, $v_0 = \smax/\langle t_a \rangle$ and $A(t_a,0)$ corresponds to the steady state solution at $v\rightarrow \infty$ (details in the text). Each curve corresponds to a different value of acceleration, which was allowed to act until $v=0$.  (b) Friction force evolution during subsequent loading for the $S$ distributions in (a) found by solving equation~\eqref{eq:mu_x_fiber_bundle} with $\nu_A(0) = \mu_{s,\text{max}}/2$ numerically. The static friction coefficients are determined from equation~\eqref{eq:fiber_bundle_friction_coefficient} as the maxima of the curves (dots). (c) Static friction coefficient \textit{vs} the corresponding deceleration underlying the stretching distribution in (a). (d) Same type of data as (c), but solved numerically, for other choices of $\Theta(\ta)$. The initial condition is the steady state solution at the initial velocity, $v_0$. As long as $v_0$ is sufficiently large, the results are independent of $v_0$ (\figurename~\ref{fig:v_0_independence}). Note that the constants in the $\Theta$ expressions determine the characteristic timescale $\langle t_a \rangle$ and therefore cancel out when dimensionless axes are used.
\label{fig:mesoscopic_static_friction_coefficient}}
\end{figure}

\figurename~\ref{fig:mesoscopic_static_friction_coefficient} (a) shows $S(s,t \gg t_\text{stop})$ scaled with the characteristic time and length of the system: $\smax S$, $\frac{s}{\smax}$, $\frac{v_0 \langle t_a \rangle}{\smax}$ and $\frac{a \langle t_a \rangle^2}{\smax}$. Each curve corresponds to a particular value of $a$. The lower the acceleration, the larger the displacement, and hence the wider is $S(s)$. For the limiting $a$, which gives $\Delta x_\text{stop} = \smax$, $S(s)$ extends from $0$ to $\smax$. Larger accelerations bring the block to rest in a shorter time, increasing $C$. In the limit $a\rightarrow\infty$, $C\rightarrow 1$ and $S(s)\rightarrow \delta(s)$.

Knowing $S$ we proceed to the next part of our argument, which is to find the static friction coefficient that results from $S$ the next time motion is triggered. We need to assume a force law in the pinned state and then solve equation~\eqref{eq:mu_x_fiber_bundle}. We use a linear elastic law, $\nu_S(s) = ks$, where $k$ has dimension 1/length. Inserting this friction law and equation~\eqref{eq:S_after_stopping} in equation~\eqref{eq:mu_x_fiber_bundle} yields an expression that has no analytical solution, but it is straightforward to perform the calculation numerically. For all calculations we used $\nu_A(0) = \mu_{s,\text{max}}/2$. The loading curves corresponding to the different $S$ are plotted in \figurename~\ref{fig:mesoscopic_static_friction_coefficient} (b). Wider $S$ result in loading curves that fall off for smaller $\Delta x$ and have lower maxima. In \figurename~\ref{fig:mesoscopic_static_friction_coefficient} (c) we show the macroscopic static friction coefficient that results from each $S$ as a function of the deceleration $a$ that gave rise to that $S$. Increasing the amplitude of $a$ makes $S$ narrower and increases $\mu_s$.

\begin{figure}
\includegraphics[width=.49\textwidth]{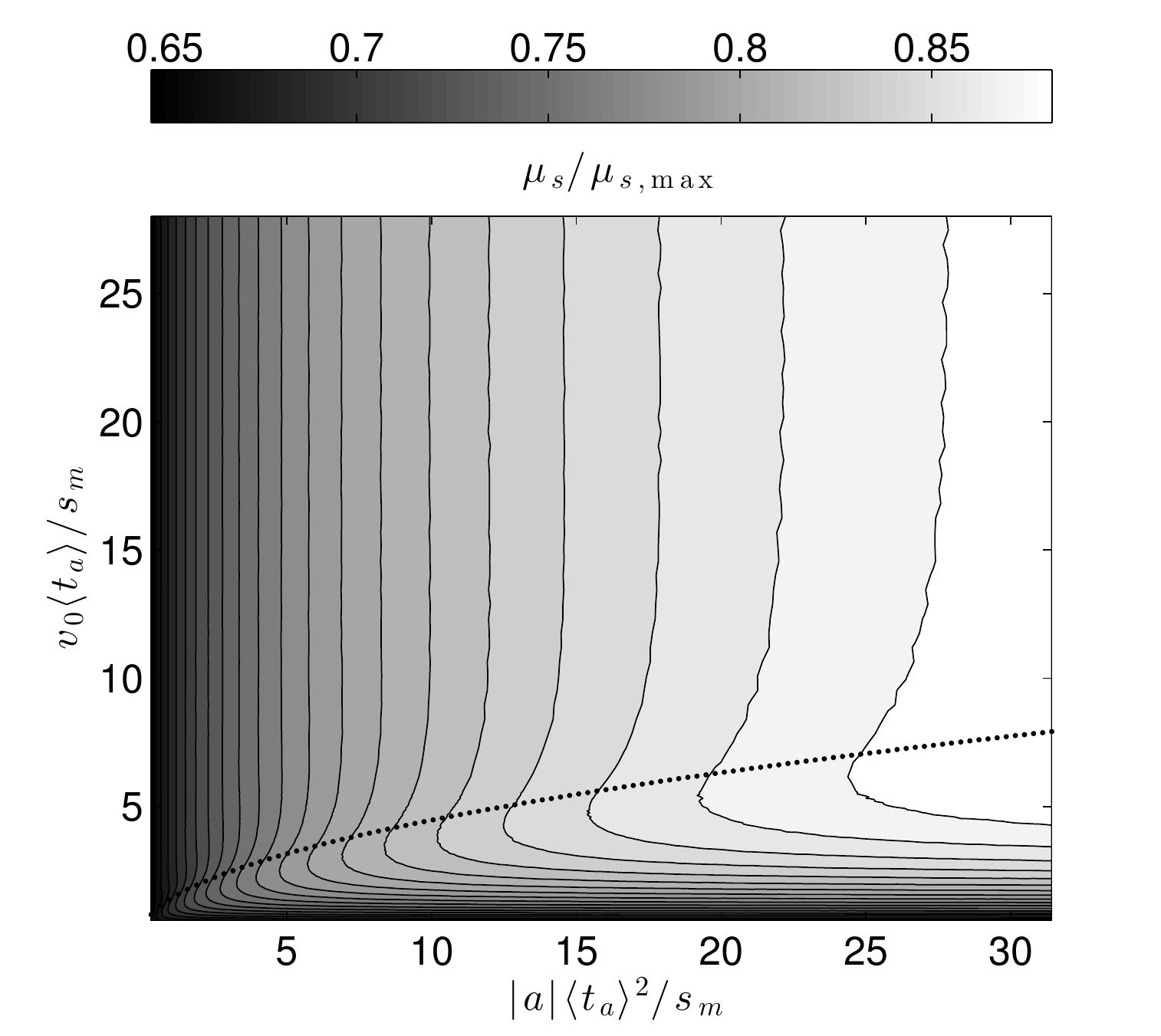}
\caption{
Numerical solution of the static friction coefficient (increasing from black to white) as a function of constant deceleration, $a$, and the initial velocity, $v_0$. The static friction coefficient is determined from equation~\eqref{eq:fiber_bundle_friction_coefficient} with $\nu_A(0) = \mu_{s,\text{max}}/2$. The initial condition is the steady state distributions at velocity $v_0$. The black dotted line is the line where total displacement during cessation, $\Delta x$, equals $\smax$. As long as $v_0$ is sufficiently high (the total displacement is sufficiently larger than $\smax$), the results in \figurename~\ref{fig:mesoscopic_static_friction_coefficient} are $v_0$ independent.  When the lines are vertical, the static friction coefficient only depends on $a$, and not $v_0$. In this figure, $\Theta = \text{const} \times t_a$. Similar results are found for $\Theta = \text{const}$ and $\Theta = e^{\text{const} \times t_a}$.
\label{fig:v_0_independence}}
\end{figure}

Since the assumptions made in the analytical case were quite strict, we also solve equation \eqref{eq:S} and \eqref{eq:A} numerically. This allows us to use as initial conditions for $S$ and $A$ the steady state solution at given initial velocity $v_0$ and to lift the restriction that the block comes to rest within $\smax$. The results are fully consistent with the results in \figurename~\ref{fig:mesoscopic_static_friction_coefficient} (a)-(c), but cover a wider range of the behaviour, in particular, lower amplitudes can be used for $a$. Also, we can check the robustness to changes in $\Theta$. The results are shown in \figurename~\ref{fig:mesoscopic_static_friction_coefficient} (d). They demonstrate the same trend as the analytical solution, and the results are not sensitive to the choice of time-dependence in $\Theta$. 

In addition, the initial velocity does not matter as long as the total displacement is sufficiently larger than the stretching threshold $\smax$. This is demonstrated in \figurename~\ref{fig:v_0_independence}. It shows the macroscopic static friction coefficient as a function of both the initial velocity $v_0$ and the acceleration $a$. When the total displacement is larger than $\smax$, the results are found insensitive to the initial velocity.

The existence of a deceleration-dependent static friction threshold could in principle be tested in laboratory experiments, provided relevant deceleration values are considered. Recent friction experiments using PMMA shows motion on a time-scale of about $400$ $\mu\text{s}$ \cite{Ben-David2010slip-stick}. In our model, this time-scale is of order $\langle t_a \rangle$ and corresponds to the time when all junctions are pinned, i.e. the motion has stopped. Assuming a time scale $\langle t_a \rangle$ in the $\text{ms}$ range, and a length scale $\smax$ in the $\mu \text{m}$ range, changes in the static friction coefficient should be observed for decelerations of the order of $\text{m}/\text{s}^{2}$ if the timescale is indeed of the same nature as in our model. Note that the effect is expected to be insensitive to the particular choice of $\Theta$, i.e the particular physical origin of the time-scale $\langle t_a \rangle$.

\subsection{Slow slip \label{sec:slow_slip}}

When a system is loaded by a force or compliant driving device rather than having its velocity prescribed, slip motion can occur at a range of speeds. Here we focus on slow slip: motion that occurs at speeds much lower than the speeds typically occuring in full sliding. Slow slip was observed in a variety of frictional systems \cite{Ohnaka1990constitutive, Heslot1994creep, Hirose2005repeating, Ben-David2010slip-stick,  Peng2010integrated, Das2011long-period, zoback2012importance,Yang2008dynamics}. It is thought to be important in geology \cite{Ohnaka1990constitutive,Hirose2005repeating,Peng2010integrated} and engineering \cite{Das2011long-period,zoback2012importance}, and we recently argued \cite{Tromborg2014slow} that it can be responsible for the slow fronts observed in laboratory friction experiments \cite{Rubinstein2004detachment}.

Consider a block obeying the friction law in equation~\eqref{eq:total_friction} being pushed by a compliant driver. Assume that the block is at or nearly at rest, for example having come to a stop after a full sliding event. We will show that slow slip is predicted in the model if 1) the flow of probability from $A$ to $S$ is associated with a drop in the friction force or 2) $\nu_A(t_a)$ is time weakening. Models corresponding to both cases were considered in \cite{Tromborg2014slow}. In our model, the origin of slow slip is the decay in friction force from slipping junctions over a time-scale $\langle t_a \rangle$. As a consequence, a slider that would tend to stop is actually continuously brought into (slow) motion by the microscopic dynamics of the junctions.

We calculate the slow slip velocity from the change in force on the macroscopic block as a function of time, which is
\begin{align}
\frac{\text{d} \nu}{\dt} = \frac{\text{d}}{\dt} \left ( \int_0^\infty \nu_A(t_a) A(t_a,t) \,\dt_a  + \int_{-\infty}^\infty \nu_S(s) S(s,t) \,\ds \right ),
\end{align}
where the first term is explicitly time-dependent, and the second term accounts for the elastic forces of the junctions. Note that this is valid if the slider is considered rigid. If the elasticity of the slider is not negligible (e.g. it is modeled as a chain of rigid mesoscopic blocks as in \cite{Braun2009dynamics, Tromborg2011transition, Amundsen20121D,Capozza2011stabilizing,Capozza2012static,Tromborg2014slow}) it has to be accounted for here. If we assume quasi-static motion of the slider, the reduction in force from the $A$ term is balanced by an equal increase in the elastic forces. The relative internal to interfacial stiffnesses determine how much is taken up by the pinned junctions and by the driving mechanism. Let us consider a soft driving mechanism, so that over the range of displacement relevant here, the driving force is constant. Then, in the quasi-static case, the change in force from the slipping junctions is balanced by a corresponding change in force from the pinned junctions, and
\begin{align}
0 = \frac{\text{d}}{\dt} \left ( \int_0^\infty \nu_A(t_a) A(t_a,t)\, \dt_a +  \int_{-\infty}^\infty \nu_S(s) S(s,t) \,\ds \right ).\label{eq:0=dnu/dt}
\end{align}

Equation~\eqref{eq:0=dnu/dt} is a general statement. To proceed with more specific calculations, let us consider an example where $\nu_A$ is constant and $\nu_S = ks$. It is possible to bring the time derivative inside the integrals, apply the definition of the derivative ($\text{d}A(\ta,t)/\dt = \left (A(\ta,t+\dt)-A(\ta,t)\right )/\dt$ ), insert equations~\eqref{eq:S} and \eqref{eq:A}, manipulate the resulting expressions and use equation~\eqref{eq:PoutofAdefined} to simplify the answer. However, the calculation is somewhat involved and tends to hide the simplicity of our argument. It is also unnecessary, as the simple case we have chosen allows us to write down the result directly from the following physical interpretation.

As stated above, the change in friction force from the slipping junctions is balanced by the change in elastic forces, which in this case means the pinned junctions only, as we have assumed that the driving force is constant. The change in force from the slipping junctions over a time step $\dt$ is $\nu_AP_\text{out of A}(t)$, where $P_\text{out of A}(t)$ is the probability that leaves $A$ during the time step. Assuming no pinned junctions break, the change in force from the pinned junctions is just $\dx\, kP_S(t)$, as they all have the same stiffness and are stretched by equally. Here $\dx$ is the displacement required to balance the forces. Combined with $\dt$, this displacement determines the slow slip velocity, $v_\text{slow slip} = \dx/\dt$. $P_S$ is the sum of its initial value $P_S(0)$ and the cumulative probability that has left $A$ (recall that by assumption no junctions are leaving $S$). Putting everything together and dividing by $\dt$ we find
\begin{align}
\begin{split}
\nu_A &\frac{P_\text{out of A} (t)}{\dt} = \\
& k\left ( P_S(0) + \int_0^t \frac{P_\text{out of A} (t')}{\dt'} \,\dt'  \right )\frac{\dx}{\dt}.
\end{split}
\end{align}
As before we can define $P_\text{out of A}/\dt = R_{A\rightarrow S}(t)$, which can be found from equation~\eqref{eq:P_OutOfA_AsAFunctionOfTime}. We can write the result as
\begin{align}
v_\text{slow slip} = \frac{\nu_A R_{A\rightarrow S}(t)}{k \left ( P_S(0) +  \int_0^t R_{A\rightarrow S}(t') \,\dt' \right ) }.
\label{eq:v_slow_slip}
\end{align}

Equation \eqref{eq:v_slow_slip} shows that slow slip exists in the model as long as $\nu_A \neq 0$. Information about the initial distribution of junctions in the slipping state is needed to calculate the rate, $R_{A \rightarrow S}$, from $A$ to $S$. As an example of applying equation~\eqref{eq:v_slow_slip}, consider $\Theta=\text{constant}$. We then have from equation \eqref{eq:expected_lifetime_in_A} that $\langle t_a \rangle = \frac{1}{\Theta}$, and from equation \eqref{eq:P_OutOfA_AsAFunctionOfTime} that $R_{A\rightarrow S}(t) = P_A(0) \Theta e^{-\Theta t}$. Inserting this in equation \eqref{eq:v_slow_slip} yields a slow slip velocity 
\begin{align}
v_\text{slow slip} & = \frac{\nu_A P_A(0) \Theta e^{-\Theta t} }{k \left ( P_S(0) +  P_A(0) (1 -e^{-\Theta t}) \right )  } \\
& = \frac{1}{\langle t_a \rangle }\frac{\nu_A P_A(0) e^{-t/\langle t_a \rangle } }{k \left ( P_S(0) +  P_A(0) (1 -e^{-t/\langle t_a \rangle}) \right )  }.
\end{align}
The dependence on 1/$\langle t_a \rangle$ is consistent with previous results \cite{Tromborg2014slow}. Note that we have chosen to neglect the contribution from bulk elasticity, so that a complete correspondence is not to be expected.

\section{Discussion - Conclusion \label{sec:summary_and_conclusion}}
\newcommand{\eg}{e.g. }
We have introduced a general framework for a frictional interface consisting of a large number of micro-junctions that can switch between a time-dependent slipping state and a displacement-dependent pinned state. The collective macroscopic state of the interface is described in terms of the evolution of two coupled probability densities, one for each state of the junctions. This general framework can be applied to a whole family of behavior laws at the microscopic scale. As a matter of fact, various models from the literature, introduced for different frictional situations, are shown to be particular subsets of this framework, \eg \cite{Farkas2005static, Braun2008modeling, Braun2010master, Braun2011dependence, Braun2009dynamics, Capozza2011stabilizing, Capozza2012static, Tromborg2014slow}. 

As the role of disorder in the properties of the pinned state of the junctions has previously been studied in detail \cite{Farkas2005static, Braun2008modeling, Braun2010master, Braun2011dependence}, we have chosen to focus on the properties of the slipping state.  To do this, we have deliberately removed the disorder in the pinned state, i.e. all junctions have the same breaking threshold. In the most generic cases in which the properties of both the pinned and slipping states would be disordered, we expect both the previously established and the present results to hold simultaneously. The detailed study of such a complete model is out of the scope of the present paper.

The time-scale that controls the duration of the microscopic slipping state has several important consequences at the macroscopic scale. Within the chosen subset of the model, we showed that the macroscopic steady sliding friction force is velocity-dependent. When the force laws of both states of the micro-junctions, as well as the probability of switching from the slipping to pinned state, are velocity independent, macroscopic friction is \emph{monotonous}; either velocity-weakening or velocity-strengthening. In contrast, as soon as the switching probability becomes velocity-dependent, the macrosocpic friction law can become \emph{non-monotonous}. We showed that a realistic description of micro-contacts at a rough interface leads to a velocity-weakening then velocity-strengthening friction, as has been repeatedly observed experimentally \cite{Chen2006velocity,bar2013velocity,Dieterich1978Time, Heslot1994creep, Baumberger2006solid}, used as a mesoscopic friction law in numerical models \cite{burridge1967model} and discussed theoretically \cite{BarSinai2012slow,bar2013velocity,BarSinai2013instabilities}.

However, most friction situations, like stick-slip motion \cite{Rubinstein2004detachment}, other friction instabilities \cite{LeRouzic2013squeal} or oscillating contacts \cite{Rigaud2013interfacial}, are out-of equilibrium regimes, possibly involving large accelerations. In these situations the steady-state friction law is irrelevant, and the coupling between the pinned and slipping populations gives rise to a \emph{history-dependent} friction behavior. We find that the macroscopic static friction coefficient at the onset of sliding is strongly influenced by the slip dynamics of the previous sliding event. In particular, the deceleration during the stopping phase of the previous sliding event controls the static friction coefficient of the next sliding event.
This effect arises through the following mechanisms: The distribution of forces among the pinned junctions controls the total force required to bring a macroscopic interface into sliding, i.e. the static friction force. The full dependency is non-trivial, but as a rule of thumb narrower distributions yield higher static friction. This relationship implies a dependence of the static friction on how the junctions are brought back into their pinned state in the preceding slip event. If the slider stops abruptly while most junctions are in the slipping state, then they will all relax and repin at low force levels, yielding a narrow distribution and thus a high static friction. Conversely, if the slider stops through a long deceleration period, the time-distributed transitions from slipping to pinned state will translate into a wide distribution of forces, as the junctions will be loaded by the moving slider after repinning, and the static friction will be low. 

Recent experiments \cite{BenDavid2011Static}  and models  \cite{Farkas2005static, Maegawa2010precursors, Scheibert2010torque, Capozza2012static, Otsuki2013systematic} have recognized that the static friction coefficient is not only controlled by the normal force and the material parameters, but is also dependent on the details of how the external loading is applied to the interface. Here we have demonstrated that it may in addition exhibit history-dependence: it varies according to how the system came to rest in the previous slip event. Note that the origin of this effect is completely different from that of the classical strengthening of interfaces due to aging. We suggest that this additional form of history-dependence of the static friction coefficient should be investigated in experiments. Our model predicts that it could be observed provided the deceleration is on the order of $\smax / \langle t_a \rangle^2$ (the characteristic length scale over the characteristic time-scale squared). In recent experiments \cite{Ben-David2010slip-stick}, we estimate this characteristic deceleration to be on the order of $\text{m}/\text{s}^{2}$.

In the above discussion, the distribution of forces in the pinned junctions appears to be a characteristic of the microscopic state of the interface which controls its future frictional behavior. This state plays a role analogous to the contact age in the classical state-and-rate friction framework \cite{Rice-Ruina-JApplMech-1983, Baumberger2006solid}. The contact age also implies a dependence of the instantaneous behavior of the interface on its history. However, the effects of the junction distribution state and the contact age state are not equivalent. To see this directly, consider an interface at rest. In the pure aging case, the strength of the interface will continuously increase in time. In our model, the strength of the interface will evolve until the last junction gets repinned. After this transient, the strength will stay constant. If the two states would be relevant simultaneously, both a transient and a continuous increase of the static friction force are to be expected. Given the generality of both states, we strongly advocate in favor of the development of improved macroscopic state-and-rate friction laws involving the standard age and the force distribution as two different internal states of the interface.

The distribution of forces from micro-junctions in the pinned state is usually hard to access experimentally, as interfacial measurements like the local stresses \cite{Scheibert2008experimental, Scheibert2009stress, BenDavid2010dynamics, BenDavid2011Static, Prevost2013probing} or the local true area of contact \cite{Rubinstein2004detachment, Maegawa2010precursors, BenDavid2010dynamics, BenDavid2011Static} are commonly made over length scales encompassing a large number of micro-junctions. The local shear stress is a direct measure of the mean value of this force distribution, but it does not carry information about its shape or width. Interestingly, the recent use of micro-structured frictional surfaces made of deterministically prepared micro-asperities \cite{Wu2010effect, Degrandi2012sliding, Broermann2013friction, Romero2014probing} may provide a way around this limitation and enable a direct test of our theory. The shear force on each individual junction can be measured through its in-plane displacement \cite{Broermann2013friction, Romero2014probing}. When the asperities are spherical, the normal force on each junction can also be measured simultaneously \cite{Romero2014probing}.

Another implication of the model is the prediction of \emph{slow slip}, which is a generic phenomenon observed in a wide variety of frictional systems \cite{Ohnaka1990constitutive, Heslot1994creep, Hirose2005repeating, Peng2010integrated, Ben-David2010slip-stick,zoback2012importance,Das2011long-period,Yang2008dynamics}. In the present model, slow slip arises from a time-dependent decrease in the friction force over a timescale on the order of the lifetime in the slipping state, and not from an external loading. The precise properties of the slow slip will depend on the microscopic force law chosen, and on the interplay between the local frictional response and the interactions with the surrounding parts of the sliding system. Note that no slow slip is expected if the only time-dependence in the junction force law is aging. Recently, slow slip has been shown to be important for rupture velocities during the onset of frictional sliding \cite{Tromborg2014slow}.

Here we have considered a single rigid slider connected to a track via a series of micro-junctions. However, elasticity is sometimes an important aspect of the dynamics of the slider, like for instance for the study of rupture front propagation across macroscopic contacts. In that case, the slider is usually divided into nodes interacting elastically, each node moving according to an effective mesoscopic law \cite{Tromborg2011transition, Kammer2012propagation, Amundsen20121D, Otsuki2013systematic} or according to the collective behavior of many junctions like here \cite{Braun2009dynamics, Capozza2011stabilizing, Capozza2012static, Tromborg2014slow}. To reduce computational costs, it is desirable to reduce the latter case to the former. By using a handful of coupled effective equations at the node scale, one could avoid the need to solve for the individual motion of a large number of micro-junctions. How to derive such effective laws from the present framework is still an open question.

\acknowledgments{
K.T. acknowledges support from VISTA - a basic research program funded by Statoil, conducted in close collaboration with The Norwegian Academy of Science and Letters.
J.S. acknowledges support by a Marie Curie FP7-Reintegration-Grants (PCIG-GA-2011-303871) within the 7th European Community Framework Programme, and by a Hubert
Curien Partnership Aurora n$^{\circ}$ 752413G.
}

\appendix

\section{Mapping between a threshold distribution and \texorpdfstring{$\Phi$}{Phi}\label{appsec:Phi_equiv_smaxDistribution}}
\newcommand{\Sb}{S_\text{m}}
In this section we show that equation~\eqref{eq:P_out_of_S},
\begin{align}
P_\text{out of S} = \dx \,\int_{-\infty}^\infty S(s,t) \Phi(s)\, \ds,
\end{align}
is mathematically equivalent to having a distribution of junction strengths, as used in Braun and Peyrard \cite{Braun2008modeling, Braun2010master}. We start by going from a distribution of strengths to $\Phi$, then we go from $\Phi$ and back, before we end with a worked example. The time equivalent of the mapping we derive here applies between $\Theta$ and a distribution of lifetimes in the slipping state, for which we give the formulae in appendix~\ref{appsec:Theta_equiv_tau}.

Let the junction strength be described by the distribution $\Sb$. Let a fraction of junctions $P_S^0$ with this strength distribution have stretching $s=0$ at some time $t$. We will follow these junctions as they stretch and break. At stretching $s$ the original probability $P_S^0$ has been reduced to
\begin{align}
P_S^s = P_S^0\int_s^\infty\Sb(s')\,\ds',
\label{eq:P_S^sFromSb}
\end{align}
because the junctions with a stretching threshold below $s$ have been broken. Within the next displacement $\dx$ the probability that leaves $S$ is 
\begin{align}
P_\text{out of S} &= P_S^0\int_s^{s+\dx}\Sb(s')\,\ds' \approx P_S^0 \Sb(s) \,\dx,
\end{align}
if $\dx \rightarrow 0$. We must find the same value when we calculate $P_\text{out of S}$ from $\Phi$. The procedure is explained in \figurename~\ref{fig:general_model_description} and gives
\begin{align}
P_\text{out of S} &= P_S^s\Phi(s)\,\dx.
\end{align}
Equating the two expressions for $P_\text{out of S}$ we find
\begin{align}
\Phi(s) &= \frac{\Sb(s)}{\int_s^\infty\Sb(s')\,\ds'},
\label{eq:PhiFromSb}
\end{align}
where we note that $P_S^0$ has cancelled out. 
The inverse mapping takes $\Phi(s)$ as the starting point and $\Sb$ as the result. The derivation is the same as the one performed for time in Appendix~\ref{appsec:AInSteadyState}. The spatial equivalent to equation~\eqref{eq:Pbreak} is
\begin{align}
\Sb(s)\dx & = \dx\, \Phi(s)e^{-\int_0^s\Phi(s')\,\ds'}
\end{align}
so that
\begin{align}
\Sb(s) & = \Phi(s)e^{-\int_0^s\Phi(s')\,\ds'}.
\label{eq:SbFromPhi}
\end{align}
A worked example will serve to make the relationship between $\Phi$ and $\Sb$ clear. We take $\Phi = \beta$ (a constant), and from equation~\eqref{eq:SbFromPhi} we find 
\begin{align}
\Sb(s) = \beta e^{-\beta s}.
\end{align}
Inserting this in equation~\eqref{eq:PhiFromSb} we find $\Phi = \beta$ as expected.
To demonstrate how the mapping works, consider a fraction $P_S^0$ of junctions which has stretching $s=0$ at some time $t$. As $s$ increases the probability remaining in $S$ evolves as 
\begin{align}
\firstderiv{P_S}{s} &= -P_S\Phi = -\beta P_S,
\end{align}
from which it follows that $P_S^s = P_S^0e^{-\beta s}$. To check for consistency we calculate $P_S^s$ from equation~\eqref{eq:P_S^sFromSb} and get
\begin{align}
P_S^s
  &= P_S^0\int_s^\infty \beta e^{-\beta s'}\,\ds'\\
  &= P_S^0e^{-\beta s},
\end{align}
which is what we also found from using $\Phi$ directly.

\section{Mapping between a delay time distribution and \texorpdfstring{$\Theta$}{Theta}\label{appsec:Theta_equiv_tau}}
\begin{figure}
\includegraphics[width=.49\textwidth]{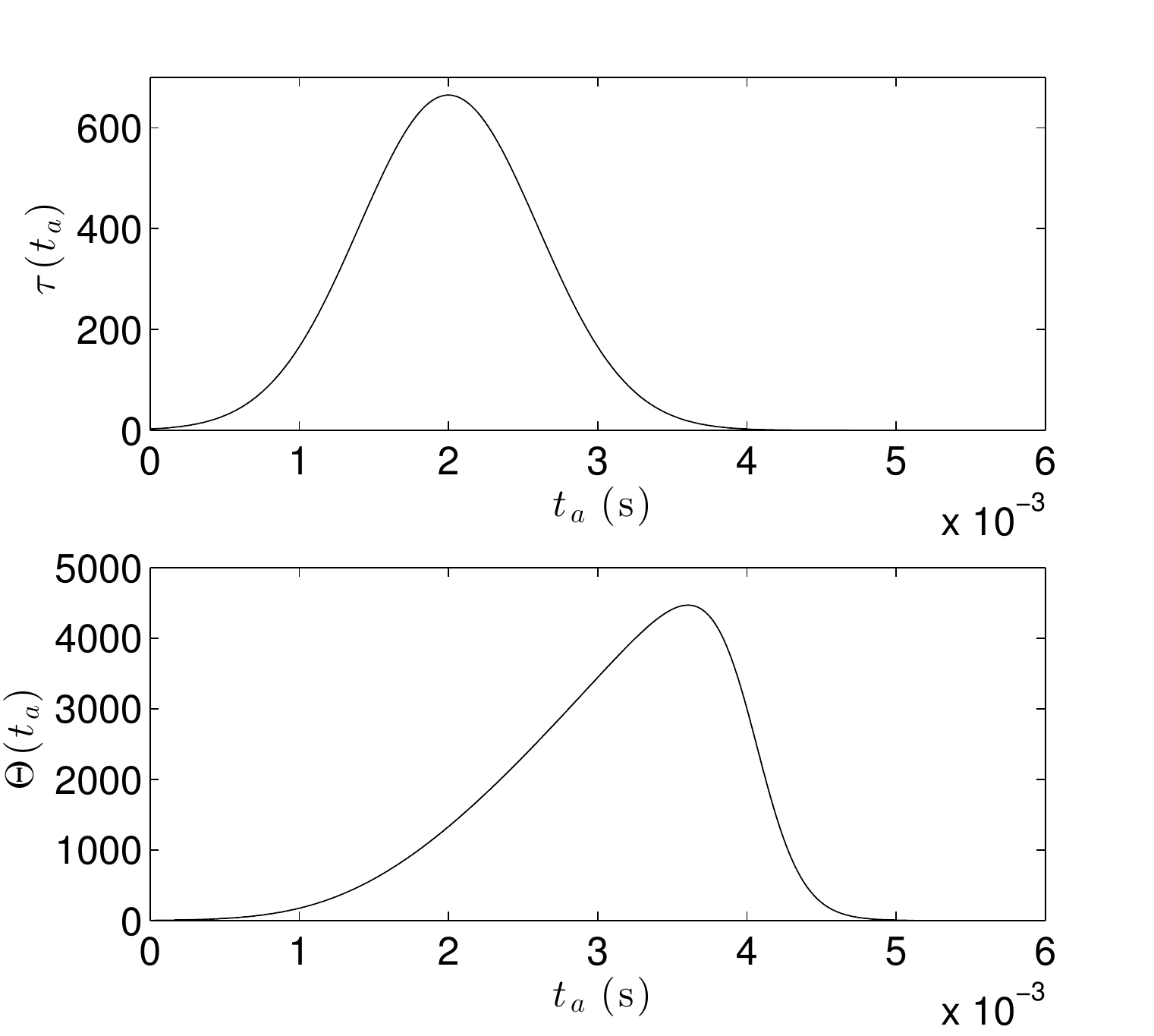}
\caption{Example of mapping from a distribution of delay times $\tau(t_a)$ (\emph{top}) to $\Theta(t_a)$ (\emph{bottom}) using equation \eqref{eq:ThetaFromTau}. Here, $\tau(t_a)$ is gaussian with mean value $2\text{ms}$ and standard deviation $0.6\text{ms}$, similar to the values used in \cite{Tromborg2014slow}.
\label{fig:map_tau_to_theta_example}}
\end{figure}

In direct analogy with equations~\eqref{eq:PhiFromSb} and \eqref{eq:SbFromPhi} we can find a mapping between a delay time distribution $\tau$ and $\Theta$. This can be useful when formulating models from the literature in our general framework. The derivation is the same as in appendix~\ref{appsec:Phi_equiv_smaxDistribution} and the corresponding results are
\begin{align}
\Theta(\ta) &= \frac{\tau(\ta)}{\int_\ta^\infty\tau(\ta')\,\dta'},\label{eq:ThetaFromTau}\\
\tau(\ta)   & = \Theta(\ta)e^{-\int_0^\ta\Theta(\ta')\,\dta'}. \label{eq:tauFromTheta}
\end{align}
In \cite{Tromborg2014slow}, we used a corrected gaussian as a distribution of delay times, $\tau$.  \figurename~\ref{fig:map_tau_to_theta_example} shows the mapping from $\tau$ to $\Theta$ for a gaussian with mean value of $2m\text{s}$ and a standard deviation of $0.6m\text{s}$. 

\section{The evolution of probability in \texorpdfstring{$A$}{A} and the steady state \texorpdfstring{$A$}{A}-distribution\label{appsec:AInSteadyState}}
This appendix includes calculations that were left out of section~\ref{sec:Steady state} in order to make the main text shorter and the main ideas more easily accessible. We detail the evolution of probability in $A$ and use this to derive $\langle\ta\rangle$ and $A_\text{steady state}(\ta)$ (equation \eqref{eq:A_steady_state_distribution}) and $\langle t_a \rangle$ (equation \eqref{eq:average t_a}).

\newcommand{\PAspike}{P_A^{\delta}}

Consider a spike of probability as it decreases in amplitude with increasing slipping time $\ta$. It enters $A$ with amplitude $\PAspike(0)$ and with $t_a = 0$. After a time $\dt$, the probability that remains in $A$ is
\begin{equation}
\PAspike(\dt) = \PAspike(0) \big(1 - \dt \,\Theta(0)\big).
\end{equation}
After one more time step the probability is
\begin{equation}
\PAspike(2\,\dt)= \PAspike(0)\big(1 - \dt\,\Theta(0) \big) \big( 1- \dt\, \Theta(\dt) \big),
\end{equation}
and after $n$ time steps, the probability is
\begin{equation}
\PAspike(n\,\dt) = \PAspike(0)\prod_{i=0}^{n-1} (1 - \dt\, \Theta(i\dt)).
\end{equation}
We find the difference between two time steps:
\begin{align}
\begin{split}
& \PAspike(n\,\dt) -\PAspike((n-1)\, \dt) \\
& = \PAspike(0) \prod_{i=0}^{n-1} \big(1 - \dt\, \Theta(i\dt)\big) - \PAspike(0) \prod_{i=0}^{n-2} \big(1 - \dt \,\Theta(i\dt)\big)\\
& = -\dt \Theta((n-1)\, \dt) \PAspike(0) \prod_{i=0}^{n-2} \big(1 - \dt\, \Theta(i\,\dt)\big) \\
& = -\dt \Theta((n-1)\, \dt) \PAspike((n-1)\, \dt) \\
\end{split}
\end{align}
\begin{equation}
\frac{\PAspike(n\,\dt) -\PAspike((n-1) \,\dt)}{\dt}= -\Theta((n-1)\, \dt) \PAspike((n-1) \,\dt).
\end{equation}
Now let $\dt \rightarrow 0$, and set $(n-1)\dt = t_a$. We find
\begin{equation}
\frac{\text{d} \PAspike(t_a)}{\dt_a}= -\Theta(t_a) \PAspike(t_a).
\label{eq:PAspike_diff_equation}
\end{equation}
This first order ordinary differential equation is to be solved with the initial condition $\PAspike(t_a = 0) = \PAspike(0)$. The equation is separable and the solution is 
\begin{equation}
\PAspike(t_a) = \PAspike(0)   e^{-\int_{0}^{t_a} \Theta(t_a')\, \dt_a'}.\label{eq:PAspike_evolution}
\end{equation}

We can use this result to calculate the average time $\langle t_a \rangle$ a junction spends in the slipping state. The probability of breaking after a time $t_a$ during a time step $\dt$ is
\begin{align}
\begin{split}
P_\text{break}(t_a) & = \PAspike(t_a)-\PAspike(t_a+\,\dt)\\
& = \dt \Theta(t_a) \PAspike(t_a)\\
& = \dt \Theta(t_a) \PAspike(0)e^{-\int_0^{t_a} \Theta(t_a')\, \dt_a'}.\label{eq:Pbreak}
\end{split}
\end{align}
The average time spent in $A$ is then
\begin{equation}
\langle t_a \rangle = \int_0^\infty t_a \Theta(t_a)e^{-\int_0^{t_a} \Theta(t_a') \,\dt_a'} \,\dt_a.
\end{equation}
This is the expectation value of $t_a$ in the distribution of delay times, $\tau$, which has a direct mapping from $\Theta$ given in equation \eqref{eq:tauFromTheta}.
$\Theta(t_a)$ is positive, and so $t_a e^{-\int_0^{t_a} \Theta(t_a') \dt_a'} \rightarrow 0$ when $t_a \rightarrow \infty$. Using Leibniz' integral rule to show that $e^{-\int_0^{t_a} \Theta(t_a') \,\dt_a'}$ is an antiderivative of $\Theta(t_a)e^{-\int_0^{t_a} \Theta(t_a') \,\dt_a'}$ we perform an integration by parts in which the surface term vanishes, to obtain
\begin{equation}
\langle t_a \rangle = \int_0^\infty e^{-\int_0^{t_a} \Theta(t_a')\, \dt_a'} \,\dt_a.\label{eq:average t_a}
\end{equation}

We now turn to the calculation of $A(\ta)$. In the steady state, probability is continuously being transferred from $S$ to $A$. It appears in $A$ with amplitude $P_\text{out of S}/\dt$ (equation~\eqref{eq:A}) and leaves $A$ according to the function $\PAspike(\ta)$ that we have already calculated. In mathematical form: $A_\text{steady state}(\ta)=\PAspike(\ta)$ when $\PAspike(0)=P_\text{out of S}/\dt$. At constant velocity, we have
\begin{equation}
P_\text{out of S} = S(\smax) \dx =  \frac{\dx}{\smax+v\langle t_a \rangle},
\end{equation}
which means that
\begin{align}
A_\text{steady state} (t_a)
  &= \frac{\dx}{\dt} \frac{1}{\smax+v\langle t_a \rangle}  e^{-\int_0^{t_a} \Theta(t_a') \,\dt_a'}\\
  &= \frac{v e^{-\int_0^{t_a} \Theta(t_a') \,\dt_a'}}{\smax+v\langle t_a \rangle}  .
  \label{eq:A_steady_state_distribution}
\end{align}
We verify that this is consistent with equation~\eqref{eq:PA} for the net probability residing in $A$ by integrating over all $\ta$ to find
\begin{align}
P_A = \int_0^\infty A(\ta)\,\dta = \PA,
\end{align}
where we used equation~\eqref{eq:average t_a}.

\section{Steady state friction coefficient, velocity dependent \texorpdfstring{$\Theta$}{Theta} \label{app:steady_state_vel_dep_theta}}
This appendix contains supplementary equations that build on and analyze equation~\eqref{eq:steady_state_friction_coefficient_velocity_dependent_theta_dimensionless}.
The behavior of the steady state friction coefficient with velocity becomes more evident if we take the derivative with respect to $\hat v$:
\begin{align}
&\frac{\partial}{\partial \hat v}\mu_\text{steady state} (\hat v) = \notag\\
& \frac{2\nu_0 \left ( (\hat \alpha + 1) s (\hat v + 1)^2 + \hat \alpha \hat v^2 \right ) - \hat k \hat s^2 (\hat \alpha + \hat v + 1)^2 }{2(\hat \alpha + \hat v + 1)^2 (\hat s \hat v + \hat s + \hat v)^2}.
\end{align}
If this derivative is negative, friction is velocity weakening, and if the derivative is positive, friction is velocity strengthening. Taking the ratio of the two terms in the numerator we conclude that $\mu_\text{steady state} (v)$ is velocity weakening if
\begin{align}
 \frac{\hat k \hat s^2 (\hat \alpha + \hat v + 1)^2}{2\nu_0\left ( (\hat \alpha + 1) s (\hat v + 1)^2 + \hat \alpha \hat v^2 \right )} > 1,
\end{align}
and velocity strengthening if
\begin{align}
\frac{\hat k \hat s^2 (\hat \alpha + \hat v + 1)^2}{2\nu_0\left ( (\hat \alpha + 1) s (\hat v + 1)^2 + \hat \alpha \hat v^2 \right )} < 1.
\end{align}
Depending on the parameter values there can also be a transition from negative derivative for small velocities to positive derivative for high velocities. The opposite is not possible because the negative term in the numerator has a $\hat v$ dependence, while the positive term has a $\hat v + \hat v^2$ dependence. (All the parameters take positive values). In the weakening-strengthening case we can locate the minimum of the friction law by solving for the transition velocity at which the derivative is zero. We find
\begin{widetext}
\begin{align}
\hat v_\text{transition} =  \frac{  \pm \sqrt{2} \sqrt{\hat \alpha^3 \hat k \nu_0 \hat s^3+\hat \alpha^3 \hat k \nu_0 \hat s^2+\hat \alpha^2 \hat k \nu_0 \hat s^3+2 \hat \alpha^2 k \nu_0 \hat s^2-2 \hat \alpha^2 \nu_0^2 \hat s+\hat \alpha \hat k \nu_0 \hat s^2-2 \hat \alpha \nu_0^2 \hat s}+\hat \alpha \hat k \hat s^2-2 \hat \alpha \nu_0 \hat s+\hat k \hat s^2-2 \nu_0 \hat s}{2 \hat \alpha \nu_0 \hat s+2 \hat \alpha \nu_0-\hat k \hat s^2+2 \nu_0 \hat s}.
\label{eq:steady_state_strenghtening_weakening_transition_velocity}
\end{align}
\end{widetext}
Three requirements for the minimum to exist follow. First, the expression must evaluate to $\hat v_\text{transition} > 0$. Second, the solution must be real, that is, the argument of the square root must be non-negative, which after factorization is seen to require
\begin{align}
\hat k \hat s \left (  \hat \alpha \hat s+\hat \alpha+1\right ) \geq 2\nu_0.
\end{align}
Third, the denominator must be different from zero for the transition to occur at a finite velocity:
\begin{align}
2\nu_0 (\hat \alpha + \hat s \hat \alpha + \hat s ) \neq \hat k \hat s^2.
\end{align}

\section{Calculating \texorpdfstring{$A(t_a,t)$}{A(t\_a,t)} after onset of slip \label{sec:A_after_onset_of_slip}}
In this appendix we calculate the distribution of slipping times $A(t_a,t)$ immediately after onset of slip. This could be used to find the evolution of the friction force during sliding initiation. Assume that we have a distribution of stretchings, $S(s,0)$, and that $P_S=1$ at $t=0$. This occurs when the slider has been at rest for a period much longer than $\langle \ta\rangle$. We start by calculating the probability that shifts from $S$ to $A$ as a function of time. Let $A_\text{initial}$ be the distribution of slip times if $\Theta = 0$. The probability that entered $A$ a time $t_a$ ago then equals the probability that left $S$ at that time because junctions reached their stretching threshold,
\begin{equation}
 A_\text{initial}(t_a,t) \dt = S(\smax,t-t_a)\, \dx.
\end{equation}
The displacement is $\dx = v(t-t_a)\, \dt$ and we find
\begin{equation}
A_\text{initial}(t_a,t)  = S(\smax,t-t_a) v(t-t_a).
\end{equation}
This is related to the initial stretching distribution $S(s,0)$ through the displacement, $\Delta x(t)$. Defining $t'=t-\ta$ we get
\begin{equation}
A_\text{initial}(t-t',t) = v(t') S(\smax - \Delta x(t'),0).
\end{equation}
The next step is to combine $A_\text{initial}$ with the decaying amplitude as probability returns to $S$. Depending on $\Theta$ the probability that enters $A$ stays there for a shorter or longer time. When it returns to $S$ it stays there until an additional displacement of $\smax$ occurs. More than one cycle of probability shift is difficult to handle analytically. We therefore restrict our calculation to the initial time period before any probability that leaves $A$ returns to $A$ again. For general $\Theta$ and non-negative $v$ this corresponds to the time interval before the displacement becomes larger than $\smax$. Then the decay in $A_\text{initial}(t-t',t)$ follows equation \eqref{eq:PAspike_evolution} so that
\begin{equation}
A(t-t',t) = A_\text{initial}(t-t',t) e^{-\int_0^{t'} \Theta(t_a) \,\dt_a},
\label{eq:A_initial_after_slip}
\end{equation}
which is valid as long as $\Theta$ does not depend on velocity. Integrating $\nu_A(t_a)$ multiplied with equation \eqref{eq:A_initial_after_slip} gives the contribution from the sliding junctions on the friction force during onset of slip.

\section{Analytical solution for the macroscopic static friction coefficient for a uniform \texorpdfstring{$S$}{S} distribution\label{appsec:staticFrictionFromUniformS}}
\newcommand{\swidth}{{s_\sigma}}
\newcommand{\sdyn}{{s_\text{A}}}
In our previous work \cite{Tromborg2014slow} we found the dependence of $\mu_s$ on the width of $S$ numerically. Here we find the analytical result for this dependence by solving equation~\eqref{eq:fiber_bundle_friction_coefficient} when $S$ is a uniform distribution of width $\swidth$. We use $\nu_S(s) = ks$ and assume that breaking is fast compared to $\langle\ta\rangle$ so that equation~\eqref{eq:mu_x_fiber_bundle} applies.

A uniform distribution of width $\swidth$ has amplitude $1/\swidth$. The friction force increases linearly as long as no junctions reach $\smax$, because all the individual contributions go up. We therefore take the point where the first junctions reach $\smax$ as our starting point. Next, we distinguish two cases based on whether the stretching of the least stretched contacts, $\smax-\swidth$, is larger than or smaller than the stretching corresponding to the dynamic force contribution, $\sdyn \equiv \nu_A(0)/k$. 

When $\smax-\swidth>\sdyn$, the total friction force goes down when junctions start breaking. To see this, consider the situation after a small displacement $\Delta x$. The junctions from the range $[\smax-\Delta x,\smax]$ have been broken and are at the force level $\nu_A(0)$. However, since $S$ is uniform the junctions from the range $[\smax-2\Delta x,\smax-\Delta x]$ have taken their place, while the junctions from the range $[\smax-\swidth,\smax-\swidth+\Delta x]$ have not been replaced. The change in the friction is then exactly the amount corresponding to shifting all the contacts in $[\smax-\swidth,\smax-\swidth+\Delta x]$ down to $\sdyn$. Since we are considering situations where $\sdyn\leq\smax-\swidth$, this reduces $\nu$. Increasing $\Delta x$ reduces $\nu$ further, and the value of $\mu_s$ is therefore attained where the first junctions break,
\begin{align}
\mu_s = \int_{\smax-\swidth}^\smax ks\frac{1}{\swidth}\,\ds = \frac{k}{2}\left(2\smax-\swidth\right), \quad \smax-\swidth>\sdyn.
\label{eq:mu_s_atfirstjunctionsbreaking}
\end{align}

Conversely, when $\smax-\swidth\leq\sdyn$, the friction force increases when junctions start breaking. This is true until the least stretched junctions are at $\sdyn$, then the friction force starts to decrease. This point is reached at displacement $\Delta x_\text{peak} = \sdyn-(\smax-\swidth)$. The force contribution from the sliding junctions is $P_A\nu_A(0) = \frac{\Delta x_\text{peak}}{\swidth}\nu_A(0)$. The contribution from the pinned junctions is $\int_\sdyn^\smax ks \frac{1}{\swidth}\ds = \frac{k}{2\swidth}\left(\smax^2-\sdyn^2\right)$, yielding the static friction coefficient
\begin{align}
\mu_s = \frac{\Delta x_\text{peak}}{\swidth}\nu_A(0) + \frac{k}{2\swidth}\left(\smax^2-\sdyn^2\right), \quad \smax-\swidth\leq\sdyn .\label{eq:mu_s_laterthanfirstjunctionsbreaking}
\end{align}

We see that in both situations $\mu_s$ is reduced as $\swidth$ increases, and that the two solutions take the same value when $\smax-\swidth=\sdyn$. Further, for the special case $\nu_A(0) = 0$, $\sdyn=0$ and $\smax-\swidth\leq\sdyn$ only when $\swidth=\smax$; the friction coefficient is $\mu_{s,\text{min}}=\frac{k\smax}{2}$ both from equation~\eqref{eq:mu_s_atfirstjunctionsbreaking} and \eqref{eq:mu_s_laterthanfirstjunctionsbreaking}. From equation~\eqref{eq:mu_s_atfirstjunctionsbreaking} we also recover $\mu_{s,\text{max}} = k\smax$ when $\swidth=0$.

In \cite{Tromborg2014slow} we compared the dependence of $\mu_s$ on the width of $S$ for different shapes of $S$. For the uniform distributions considered here the width can be defined simply as $\swidth$, but in general the standard deviation would be our preferred choice, as it can be defined for distributions of arbitrary shape.

\bibliography{Thogersen_main_text.bbl}

\end{document}